\documentclass{aa}
\usepackage{txfonts}
\usepackage{natbib}
\usepackage{footnote}
\usepackage{enumerate}

\usepackage{graphicx}

\def\dex{\,{\rm dex}}

\def\url#1{{\tt#1}}

\begin{document}

\title{The nuclear stellar disc of the Milky Way: \\  A dynamically cool and metal-rich  component possibly formed from the central molecular zone}

\author{  M. Schultheis \inst{1} 
  \and   T.K. Fritz  \inst{2,3}
  \and G. Nandakumar \inst{4,5}
  \and A. Rojas-Arriagada\inst{6,7}
  \and F. Nogueras-Lara\inst{8}
  \and  A. Feldmeier-Krause\inst{9,8}
  \and O. Gerhard\inst{10}
  \and N. Neumayer\inst{8}
  \and L.R.  Patrick \inst{11}
  \and M.A. Prieto \inst{2,3}
  \and R. Sch\"odel \inst{12}
  \and A. Mastrobuono-Battisti \inst{13}
  \and M.C. Sormani \inst{14}
}

   \institute{ Universit\'e C\^ote d'Azur, Observatoire de la C\^ote d'Azur, Laboratoire Lagrange, CNRS, Blvd de l'Observatoire, F-06304 Nice, France
 e-mail: mathias.schultheis@oca.eu
\and
Instituto de Astrof\'{i}sica de Canarias, Calle Via L\'{a}ctea s/n, E-38206 La Laguna, Tenerife, Spain
\and
Universidad de La Laguna (ULL), Departamento de Astrof\'{i}sica, E-30206 La Laguna, Tenerife, Spain
\and
 Research School of Astronomy \& Astrophysics, Australian National University, ACT 2611, Australia
 \and
 Centre of Excellence for Astrophysics in Three Dimensions (ASTRO-3D), Australia
 \and
 Instituto de Astrof\'isica, Facultad de F\'isica, Pontificia, Universidad Cat\'olica de Chile, Av. Vicu\~na Mackenna 4860, Santiago 8970117,  Chile
 \and
 Millennium Institute of Astrophysics, Av. Vicu\~na Mackenna 4860, Macul, 
Santiago 7820436, Chile
 \and
 Max-Planck-Institut f\"ur Astronomie, K\"onigstuhl 17, 69117, Heidelberg, Germany
 \and
 The Department of Astronomy and Astrophysics, The University of Chicago, 
5640 S. Ellis Ave., Chicago, IL, 60637, USA
 \and
 Max-Planck-Institute for extragalactic Physics, Giessenbackstra{\ss}e 1, 
D-85748 Garching, Germany
 \and
 Departamento de F\'isica Aplicada, Facultad de Ciencias, Universidad de Alicante, Carretera San Vicente s/n, E-03690, San Vicente
 \and
 Instituto de Astrofisica de Andalucia (CSIC), Glorieta de la Astronomia s/n, 18008 Granada, Spain
 \and
 Department of Astronomy and Theoretical Physics, Lund Observatory, Box 43, SE--221 00, Lund, Sweden
 \and
 Universit\"at Heidelberg, Zentrum für Astronomie, Institut für 
theoretische Astrophysik, Albert-Ueberle-Str. 2, 69120 Heidelberg, Germany
}

\abstract {The  nuclear stellar disc (NSD)   is, together with the nuclear star cluster (NSC) and the central massive black hole,  one of the main 
components in the central parts of our Milky Way. However, until recently,  only a few studies of the stellar content  of the NSD  have been obtained   owing to extreme extinction and stellar crowding.}
 {We study the kinematics and global metallicities of the NSD based on the observations of  
K/M giant stars via a dedicated KMOS (VLT, ESO) spectroscopic survey.}
 { We traced radial velocities and metallicities, which were derived based on spectral indices (Na I and CO) along the NSD, and compared those with a Galactic bulge sample of APOGEE (DR16) and data from the NSC.}
{We find that the  metallicity distribution function and  the fraction of 
metal-rich and metal-poor stars in the NSD are  different from the corresponding distributions and ratios of the NSC   and the Galactic bulge. By tracing the velocity dispersion as a function of metallicity, we clearly see  that the NSD is kinematically cool and that the velocity dispersion decreases with increasing metallicity  contrary to the  inner bulge sample of APOGEE ($\rm |b| < 4^{o}$). Using molecular gas tracers ($\rm H_{2}CO$, CO(4-3))  of the  central molecular zone (CMZ),  we find an astonishing agreement between the gas rotation and the rotation of the metal-rich population. This agreement indicates that the metal-rich stars could have formed from gas 
in the CMZ. On the other hand, the metal-poor stars show a much slower  rotation profile with signs of counter-rotation, thereby indicating that these stars have a different origin.  }
{Coupling kinematics with global metallicities, our results demonstrate that the NSD is chemically and kinematically distinct with respect to the inner bulge, which indicates a different   formation scenario.}
\keywords{Galaxy: nucleus, structure, stellar content -- stars: fundamental parameters: abundances -infrared : stars}
\titlerunning{Nuclear stellar disc of the Milky way}
\maketitle


\section{Introduction}
The nuclear stellar disc (NSD) is a dense stellar structure in the centre 
of our Milky Way. It is embedded in the bulge and surrounds the nuclear star cluster (NSC) with its central massive black hole (\citealt{Launhardt2002}). The NSD extends up to a radius of $\rm 1.55^{o}$ ($\rm \sim  220\,pc$) with  a scale height of $\rm \sim 0.3\,^{o} (\sim 50\,pc$, \citealt{Launhardt2002}, \citealt{nishiyama13},  \citealt{nogueras2020}, \citealt{Gallego-Cano2020}).  It contains large amount of interstellar dust (\citealt{schultheis2009}, \citealt{Schoedel2014}, \citealt{Nogueras2018}), making it necessary to study the stellar population in the infrared.

\begin{figure*}[!htbp]
  \centering
        \includegraphics[width=0.98\textwidth,angle=0]{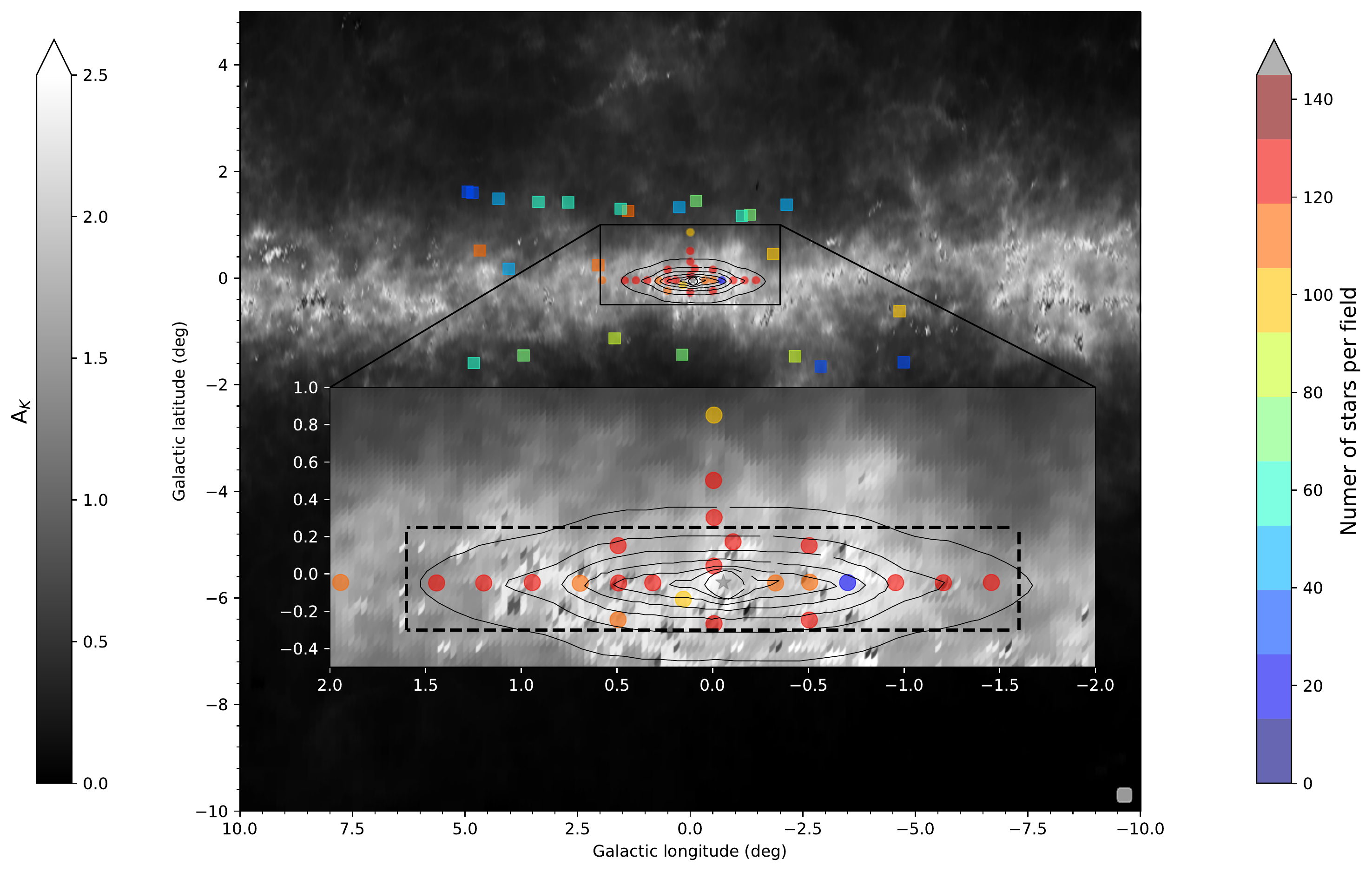}
        \caption{Location of the different KMOS fields of our sample (filled circles) superimposed on the extinction map of \citet{gonzalez2012}. The filled squares indicate the APOGEE DR16 comparison sample. The colour 
scale represents the number of objects,  while the grey scale the $\rm A_{Ks}$ value of the extinction map. The black contours show the surface brightness map of the best-fit model of the nuclear bulge component by Launhardt et al. (2002)}
        \label{Fields}
\end{figure*}

  While in  recent years studies were focussed on the NSC itself (see e.g. \citealt{Neumayer2020}, \citealt{Schoedel2020}), very little work has been done on the NSD so far. Nuclear stellar discs are also detected in extragalactic systems and are quite common in spiral  galaxies (\citealt{Pizzella2002}, \citealt{Gadotti2019}). This indicates a nuclear stellar discs have a different formation  scenario (\citealt{Pizzella2002}).

 Figer et al. (\citeyear{Figer2004}) determine, for the first time, the star formation history (SFH) in some pencil beam fields located in the NSD  
using deep HST (NICMOS) observations. These authors conclude a quasi-continuous SFH with no signs of a starburst activity.
 \citet{nogueras2020} determined the SFH of the NSD by analysing the luminosity function compared to stellar evolutionary models based on deep GALACTICNUCLEUS data  (\citealt{Paco2019a}). These authors find that the bulk of the stars in the NSD is old   and formed at least 8\,Gyr ago followed by an extended phase of quiescence and  recent star formation activity about 1\,Gyr ago, where 5\% of the mass of the NSD was formed very quickly.  With a similar technique, \citet{Schoedel2020} trace the SFH of the NSC, also finding that $\sim 80\%$ of its stars are at least about 10\,Gyr old. These authors also find 
 signs of an intermediate-age population ($\rm \sim 3\,Gyr$), but contrary to the NSD the NSC shows no signs of star formation about 1\,Gyr ago. Their results indicate different formation histories for the NSD and NSC.
  \citet{schoenrich2015} use the kinematics of APOGEE stars to trace the 
rotation velocity of the nuclear disc ($\rm \sim 120\,km/s$), suggesting that the NSD is kinematically cool with a small vertical extent of 50\,pc and a truncation radius of $\rm R \sim 150\,pc$ (see also \citealt{Gallego-Cano2020}).  In addition, these authors show that the NSD is rotating with similar velocities as the molecular gas
  in the central molecular zone (CMZ), thus indicating  that these two components could be linked together and that the stars in the NSD could originate 
from the dense CMZ gas (\citealt{Sormani2020}). With additional data from 
APOGEE DR16 (\citealt{Ahumada2019})  and SiO maser data of AGB stars in the inner Galaxy from \citet{Messineo2005}, \citet{Sormani2020} fit axisymmetric Jeans dynamical models and determine the total mass of the NSD 
 of  $M_{\rm NSD} = 6.9 \pm 2 \times 10^8 {\rm \, M_\odot}$, which is  lower than the photometric mass found by \citet{Launhardt2002}. 
  
  From a theoretical point of view, hydrodynamic simulations of Milky-Way like galaxies have shown that   the formation of the bar  triggers gas 
funnelling to the centre of the Milky Way forming a kinematically cold, rotating NSD (see e.g. \citealt{Fux1999}, \citealt{Li2015}, \citealt{Ridley2017}, \citealt{Sormani2018b}, \citealt{Sormani2018}, \citealt{Sormani2019a}, \citealt{Tress2020}). The age distribution of the stars in the NSD  can give us therefore hints about the formation epoch of the bar (\citealt{Baba2020}, \citealt{Gadotti2019}). These authors  also show that 3D velocity information  is an effective 
way to reduce the contamination of bulge stars in the NSD as stars in the 
NSD are kinematically cool in contrast to the bulge.
  While a lot of effort has been dedicated lately to obtaining precise chemical abundances   (\citealt{rich:17}, \citealt{Thorsbro2020}) and global metallicities in the NSC (\citealt{do2015,Do2018}, \citealt{Feldmeier-Krause2017,Feldmeier-Krause2020}),  very little similar work  has been done 
so far on the NSD.
\citet{Schultheis2020} use the latest APOGEE DR16 data to trace the chemistry and the kinematics of the NSD stars. The chemical abundances and the metallicity distribution function (MDF) show differences between the NSC and the NSD, suggesting a different formation scenario. However, owing to the high interstellar extinction, APOGEE could only target the brightest sources in the NSD including AGB stars and supergiant stars (\citealt{zasowski2013,zasowski2017}). Interestingly, \citet{Schultheis2020}  discovered  a relatively high number of supergiant stars, which is consistent with the SFH during the last 200--300 Myr (\citealt{nogueras2020}).

In this paper we use dedicated observations of the NSD obtained with 
the  KMOS spectrometer at the ESO VLT. The detailed survey strategy and the data reduction are described in detail by \citet{fritz2020} (paper I).  For their sample of $\sim 3000$ stars, these authors  derive radial velocities with a median velocity error of 5\,km/s. They use the NaI index and 
the  CO index in the K band  to measure global  metallicities, which were calibrated against available medium- and high-resolution metallicities.   We refer to paper I for a detailed description of the survey, instrument set-up, data reduction, and analysis. We  use global metallicities and radial velocities to trace the chemical and kinematical properties of the NSD and compare these properties with the Galactic bulge (\citealt{Ahumada2019}) and the NSC (\citealt{Feldmeier-Krause2020}).

\section{Samples}

Figure~\ref{Fields}  shows the location of our fields inside the NSD  denoted by the dashed box with colours indicating the number of objects in each KMOS field.  In addition  four fields in the Galactic bulge were observed, three along the minor axis and one in the Galactic plane. We limited our sample to a signal-to-noise ratio (S/N) better than 30 and we used only primary targets. All stars have effective temperatures cooler than 5500\,K and were selected based on the dereddened K magnitudes with $\rm 7 < K_{0} < 9.5$, which corresponds to $\rm 7.5 < K < 14$ (see also paper I).

\begin{figure}[!htbp]
  \centering
        \includegraphics[width=0.49\textwidth,angle=0]{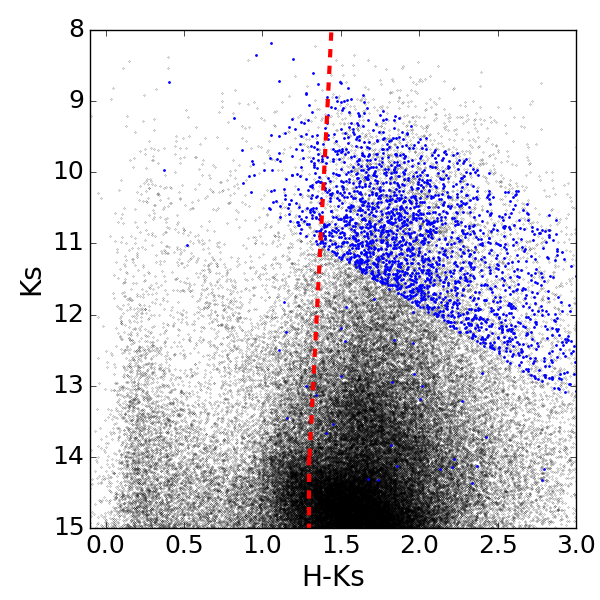}
        \caption{CMD $K_s$ vs. $H-K_s$ for the regions corresponding to the NSD. The blue points correspond to the stars in the KMOS survey. The black dots indicate stars within the same regions from the SIRIUS/IRSF GC survey (e.g. \citealt{Nagayama2003}, \citealt{Nishiyama2006}). The red dashed line shows the colour cut used to remove foreground stars from the Galactic disc and the innermost bulge following previous work (e.g. \citealt{Nogueras2018}, \citealt{Paco2019a}, \citealt{Sormani2020}).}
        \label{CMD}
\end{figure}

In total our bulge sample consists of  448 stars, while our NSD sample consists of 2118 stars. We used data from paper I, where part of the foreground population was excluded using colour cut H-Ks. To remove the remaining 
foreground sources that might affect our results, we applied more restrictive cuts following previous studies on the NSD and the innermost bulge (e.g. \citealt{Paco2018}, \citealt{Paco2019a}, \citealt{Sormani2020}). Given the high differential extinction (\citealt{nogueras2020a}) and its variation on arcsecond scales in the Galactic centre (GC, \citealt{Nogueras2018}), we applied different H-Ks cuts for each of the considered regions. 
Fig.~\ref{CMD} shows the applied colour cut for the NSD ($\rm H-Ks = max(1.3,-0.0233 \times Ks + 1.63$), where the red line shows the colour cut to remove foreground stars. For the bulge fields we used variable colour cuts for each field (3,4,5 and 28) individually allowing us to avoid disc foreground and background stars.


The overall S/N  is pretty high with the majority of our targets lying in the range $\rm 50 < S/N < 100$.  We transformed the heliocentric radial velocities to galactocentric velocities by assuming  $\rm v_{GC} = v_{rad} + 11.1*cos(l)*sin(b)+248.5*sin(l)*cos(b)+8.5*sin(b),$  where $\rm v_{rad}$ is the heliocentric radial velocity and (l,b) galactic longitude and galactic latitude. We used the metallicity calibration from paper I, which depends on the equivalent width of the CO band at 2.3$\rm \mu$m  and the Na I line ($\rm 2.21 \mu$m). The typical standard deviation compared to their reference sample
 (APOGEE, XSL, FIRE)  is about 0.32\,dex containing most  (72\%) of the stars   (see Fig.~10 of paper I).

\begin{figure}[!htbp]
  \centering
        \includegraphics[width=0.49\textwidth,angle=0]{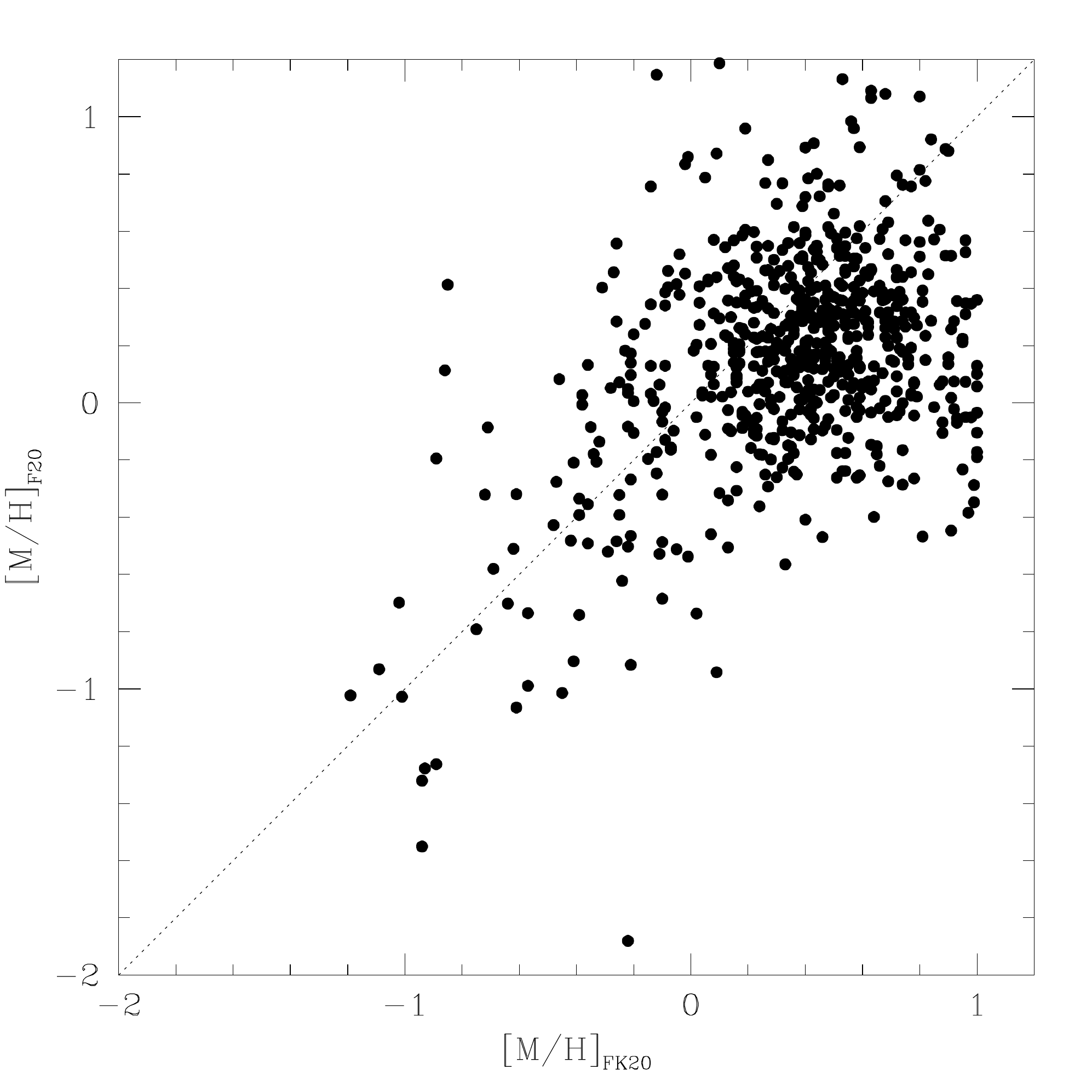}        \caption{Comparison of the metallicities of \citet{Feldmeier-Krause2020} for the NSC compared  to the metallicities derived from our indices. } 
        \label{comp_anja}
\end{figure}

As a comparison sample we took the latest APOGEE DR16 dataset (\citealt{Ahumada2019}) and constructed our bulge sample by applying a cut in the galactocentric distance $\rm R_{GC} < 3.5\,kpc$  in the same way as described in \citet{alvaro2020}. In addition, we used the KMOS sample of  K/M giants of \citet{Feldmeier-Krause2020}[FK20] located in the NSC. In order to get a consistent estimate of the metallicities, we applied our technique based on spectral indices (see paper I for  a detailed description)  to the dataset of the latter authors. We recall that  \citet{Feldmeier-Krause2017} do a full spectral fitting (STARKIT code) by using synthetic spectra to derive the stellar parameters ($\rm T_{eff}$, log\,g, and $\rm [M/H]$) by obtaining the best fit of the stellar spectrum. 
Their mean uncertainties are $\rm \sigma_{Teff} = 212\,K$, $\rm \sigma_{log\,g} = 1.0\,dex$, and $\rm \sigma_{[M/H]} = 0.26\,dex$, respectively. However, as mentioned by \citet{Feldmeier-Krause2017} the absolute metallicities are derived from synthetic spectra extending to $\rm [M/H]=1\,dex$, but only calibrated on empirical spectra for $\rm  [M/H] < 0.3\,dex$. For this reason, metallicities at $\rm [M/H]>0.3\,dex$ may be overestimated, while their systematic uncertainties may be underestimated. The metallicity measurements of Fritz et al. (2020) are calibrated using empirical spectra $\rm [M/H] <0.6\,dex$ and extrapolated beyond this value. High-resolution spectroscopic studies did not find metallicities higher than 0.6 dex in the Galactic centre  (see e.g. \citealt{rich:17}, \citealt{Thorsbro2020}). Figure~\ref{comp_anja} shows a comparison between the NSC sample by FK20 and our metallicities based on the CO and NaI index (see paper I). The median difference is 0.015\,dex with a standard deviation of 0.35\,dex, which is smaller than the uncertainties of 
each of the methods   assuring that there is no systematic bias introduced by our method. We note however that for very metal-rich stars ($\rm [M/H] > +0.5\,dex$), FK20 shows significantly  higher metallicities.

\section{Metallicity distribution function} \label{sec:GMM}

\begin{figure}[!htbp]
  \centering
        \includegraphics[width=0.49\textwidth,angle=0]{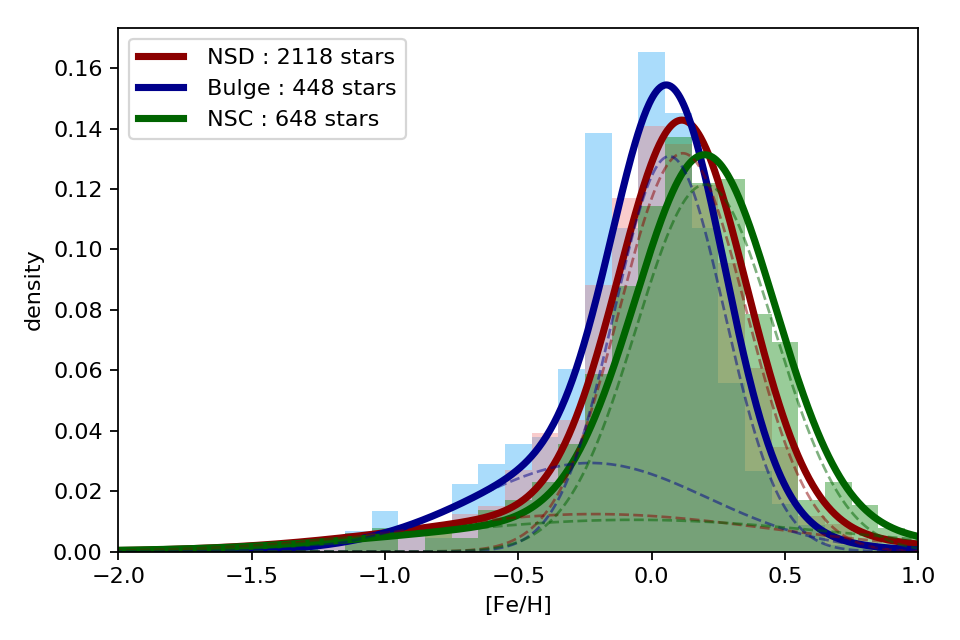}
        \caption{GMM decomposition of the  NSD sample (red), the bulge comparison sample (blue), and stars in the NSC from FK20 (green). The individual GMM  components are indicated as dashed lines, while the global  GMM 
solution by a thick solid line.} 
        \label{MDF}
\end{figure}

To study the MDF, we performed a Gaussian mixture modelling (GMM) decomposition, which is a parametric probability density function given by the weighted sum of a number of Gaussian components. The GMM parameters are estimated as those that show the best representation of the dataset density distribution structure. The expectation-maximisation algorithm determines the best parameters of a mixture model for a given number of Gaussian components.  In order to constrain the number of Gaussians, which is a priori not known, we adopted the Bayesian information criterion (BIC).  Figure~\ref{MDF} shows the results of our GMM analysis for our NSD  sample as well as our bulge and NSC sample. In all cases the GMM gives preference to a two Gaussian component solution for the three regions. We see for the three samples (NSD, bulge, and NSC) a very prominent metal-rich peak
 and a much weaker metal-poor peak. When we compare our bulge MDF with that of APOGEE (\citealt{alvaro2020}), we notice that  APOGEE finds three distinct Gaussian components with nearly constant metallicity positions (i.e. independent of Galactic latitude)   at +0.32\,dex, -0.17\,dex, and -0.66\,dex.  While the positions of these three Gaussian  components remain constant in the bulge, their relative  weights  vary and are responsible for the vertical  variations of the density shape of the MDF.  However, \citet{alvaro2020} also show  that if we use   a  smaller sample of stars (e.g. $\sim$  600) or  they have larger metallicity uncertainties, which is the case for our sample,   the three components are smoothed out to two main 
major peaks (metal-rich and metal-poor). This is similar to what we see in our MDF 
(see Fig.~\ref{MDF}),  where our  metal-poor peak is  less prominent than seen in APOGEE. 
 
 We note the dominant metal-rich peak  (89\% of the weight) of the 
NSC sample with a peak position of the metal-rich peak  at 0.22\,dex (see 
Tab.~\ref{Tab1}), which is 0.2 dex more metal-rich than the bulge sample and 0.1 dex compared to the NSD. We clearly see three distinct populations, 
and in Tab.~\ref{Tab1} we show the corresponding values based on our GMM analysis such as the peak position and their dispersion as well as the weights. A two sample Kolmogorov-Smirnov  (KS)  test confirms this hypothesis where the p-values are extremely small ($\rm  \sim 10^{-6}$),  indicating that we can reject the null hypothesis, and therefore the three samples 
do not come from the same population. We also performed a k-sample Anderson-Darling test, which confirms the findings of the KS test (p-value is $\sim 0.001$).

\begin{table}
\caption{Properties of the GMM decompositions for the Bulge, NSD and NSC. Columns: (1) Metal-rich peak metallicity in GMM, (2) metal-poor peak metallicity in GMM, (3) and (4)  width of MR and MP peak, (5) relative weight of MR peak (6) relative weight of MP peak. }  
\begin{tabular}{ccccccc}
  &MR & MP & $\rm \sigma_{MR }$&$\rm \sigma_{MP }$&$\rm W_{MR}$&$\rm W_{MP}$\\
  \hline
 Bulge &0.04&-0.55&0.21&0.33&0.82&0.18\\
 NSD &0.12&-0.22&0.24&0.64&0.78&0.22\\
 NSC&0.22&-0.41&0.35&0.18&0.89&0.11\\
\end{tabular}
\label{Tab1}
\end{table}

\begin{figure}[!htbp]
  \centering
        \includegraphics[width=0.49\textwidth,angle=0]{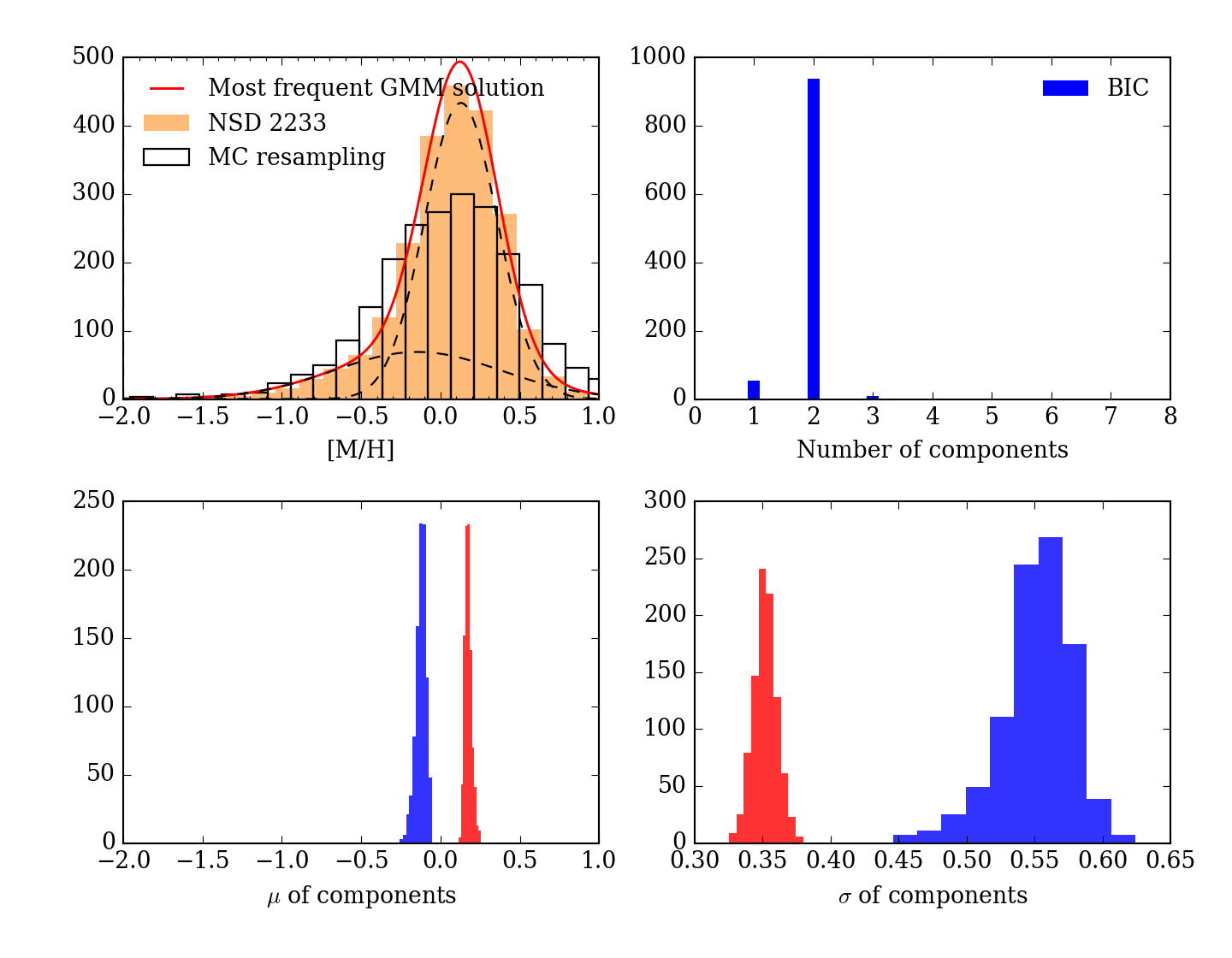}
        \caption{Monte Carlo simulations of the GMM decomposition of the  
NSD sample (orange). The upper left panel shows the individual Gaussian components and the most frequent GMM solution (red). The upper right panel shows the frequency distribution of the optimal number of Gaussian components  found on 600 MC resamplings of the observed MDF. The lower panels  show the frequency distributions  of  the centroids (left) and widths (right) of each Gaussian component.} 
        \label{MC}
\end{figure}

To evaluate the robustness of these solutions, we performed a Monte Carlo (MC) resampling by drawing 1000 MC resamplings of
the individual metallicities for each star by assuming a typical Gaussian 
error of 0.3 dex in $\rm [Fe/H]$ which is the error based on the calibration (see paper I). This is a conservative error since the median error is 
about 0.12\,dex based on the S/N ratio.
 Figure~\ref{MC} shows a typical example for the NSD, where the upper left 
panel shows the original dataset and  a random perturbation of the original distribution.  The most frequent GMM solution is also indicated together with their individual Gaussian components. The GMM parametrisation was calculated 1000 times considering mixtures with N=1-8 Gaussian components and using the BIC to identify the best model in 
each case.  As shown on the right upper panel,  a bimodal solution was found to be in 95\% cases the optimal solution. The lower panels in Fig.~\ref{MC} show the distributions of the centroids and the widths of the individual Gaussian components as fitted to each of the 1000 MC resamplings. The narrowness of these distributions indicates that the individual Gaussian  component parameters are robust over the MC resampling and that in general the observed density distribution can be fairly well described by a mixture of two Gaussians.

\begin{figure}[!htbp]
  \centering
        \includegraphics[width=0.49\textwidth,angle=0]{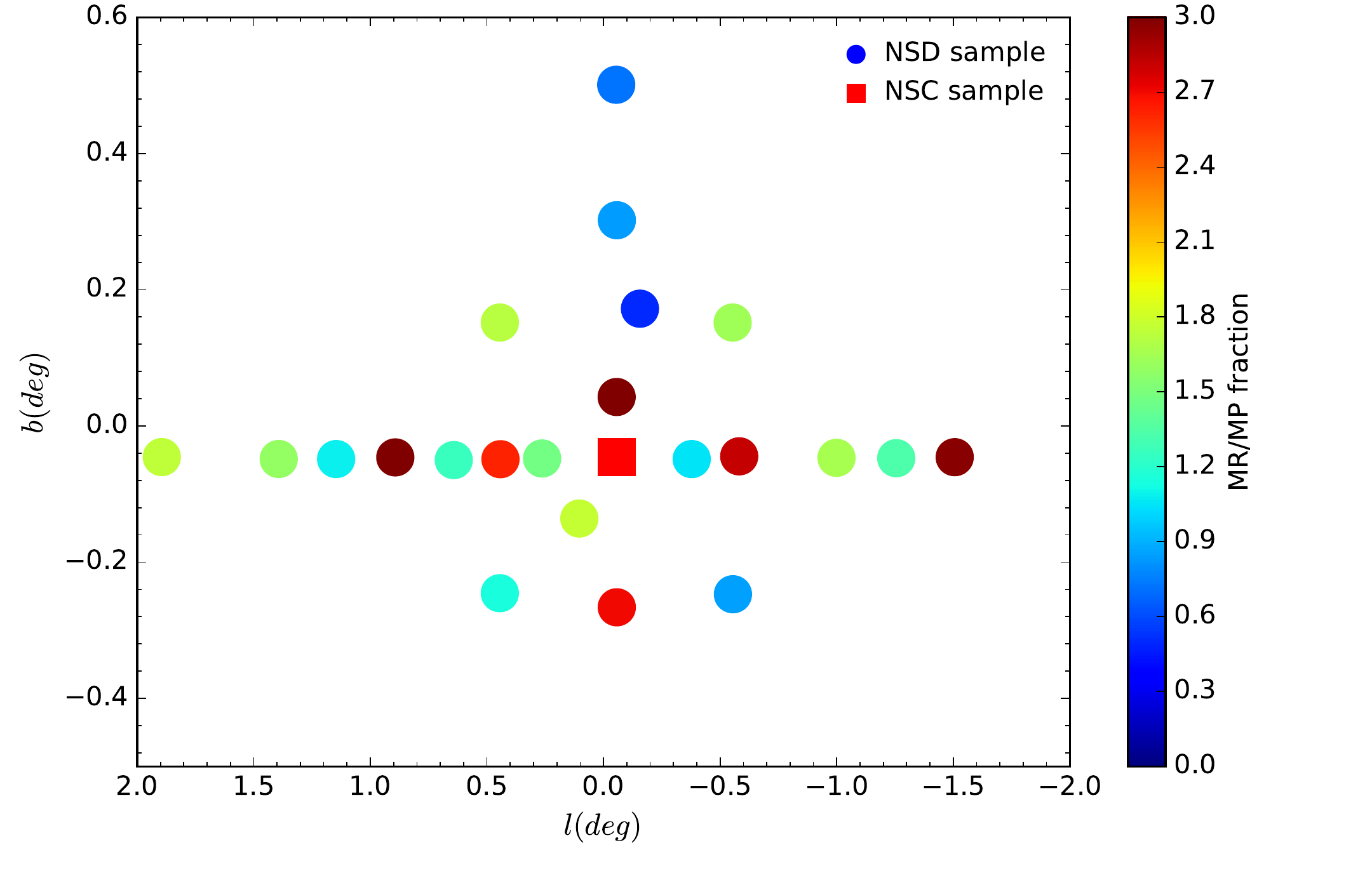}
        \caption{Fraction of MR  ($\rm [Fe/H] > 0$) to MP  ($\rm [Fe/H] < 
0$) stars in the full sampled region. Each point denotes one KMOS field.} 

        \label{fraction}
\end{figure}

We   split our samples into two large metallicity bins, a metal-rich sample 
(MR) with $\rm [Fe/H] > 0$, and a metal-poor sample (MP)  with $\rm [Fe/H] < 0$, similar to how it was done for the GIBS data by \citet{zoccali2017} or \citet{schultheis2019} for the inner Galactic bulge, including 
the Galactic centre. 
In Fig.~\ref{fraction}, we trace the fraction of MR to MP stars as  a function of Galactic longitude and latitude.  We note that at higher latitudes the MP fraction  increases, while for the majority of the fields  in the NSD the MR fraction is rather high (see also Sect. 3.1). Interestingly, we find a rather large variation of the MR/MP fraction inside the NSD with a variation of  up to nearly a factor of 10. 
We do not find any hint of an asymmetric distribution in the  MR/MP ratio 
 in the NSD as was found recently on much smaller scales  in the NSC by FK20.  In particular, we do not see any difference in  the MDF  between 
negative and positive Galactic longitudes; a KS test results in a high p-value of 0.5. We conclude therefore that we are dealing with a symmetric behaviour in the chemistry of the NSD. 


\begin{figure}[!htbp]
  \centering
        \includegraphics[width=0.40\textwidth,angle=0]{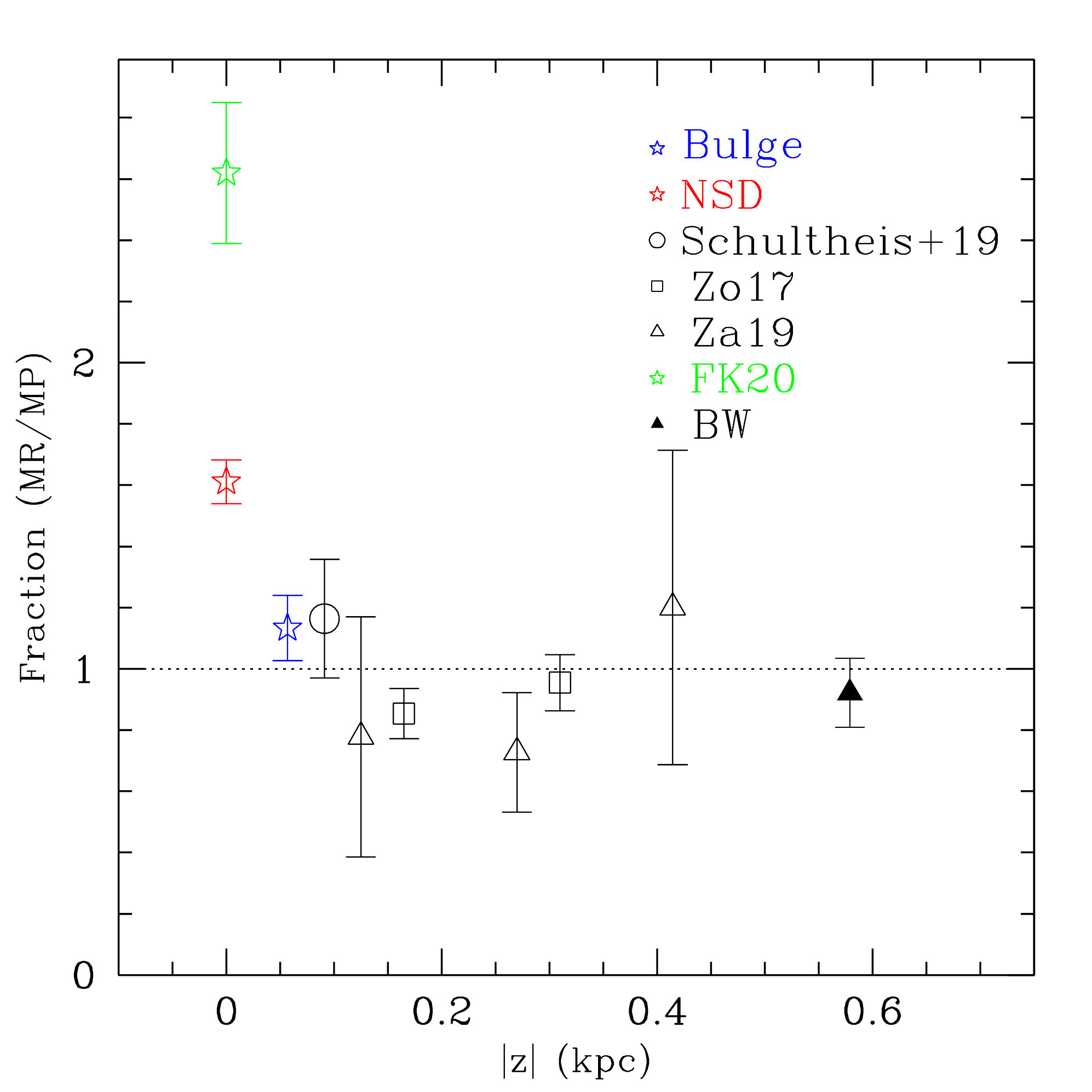}
        \caption{Fraction of MR to MP stars as a function of the scale height z. The green symbol  shows the NSC comparison sample of FK20.} 
        \label{fractionz}
\end{figure}

As shown in Fig.~\ref{MDF}, the Gaussian decomposition of the MDF in the NSD, the bulge comparison field, and the NSC  seem to show  distinct Gaussian components.  Figure \ref{fractionz} shows a compilation of the inner bulge data from different literature works taken from \citet{schultheis2019} updated with our work, the NSD, the bulge field, and the NSC (FK20). As already discussed in \citet{schultheis2019}, the fields in the bulge show a nearly constant fraction of about 50\% of MR and MP stars. Our bulge comparison  field perfectly agrees with this number. The  corresponding error bars were calculated as $\rm {\Delta~f= f\times\sqrt{(1+f)/MR}}$, where f is the number ratio of  MR to MP stars and MR  the number of MR stars. We see that for the NSD the fraction of MR to MP stars rises up  to 1.5, while  for the NSC we see an extremely high fraction  of metal-rich stars  as already pointed out by \citet{schultheis2019}, \citet{do2015}, \citet{Feldmeier-Krause2017}, and FK20.  We note that to be consistent we applied the same method to derive metallicities as in our NSD  and bulge samples for the NSC sample.
Figure~\ref{fractionz} affirms our conclusion that the three components are chemically distinct with a steadily increasing fraction of MR stars for the bulge, the NSD, and the NSC, respectively.

\begin{figure}[!htbp]
  \centering
        \includegraphics[width=0.49\textwidth,angle=0]{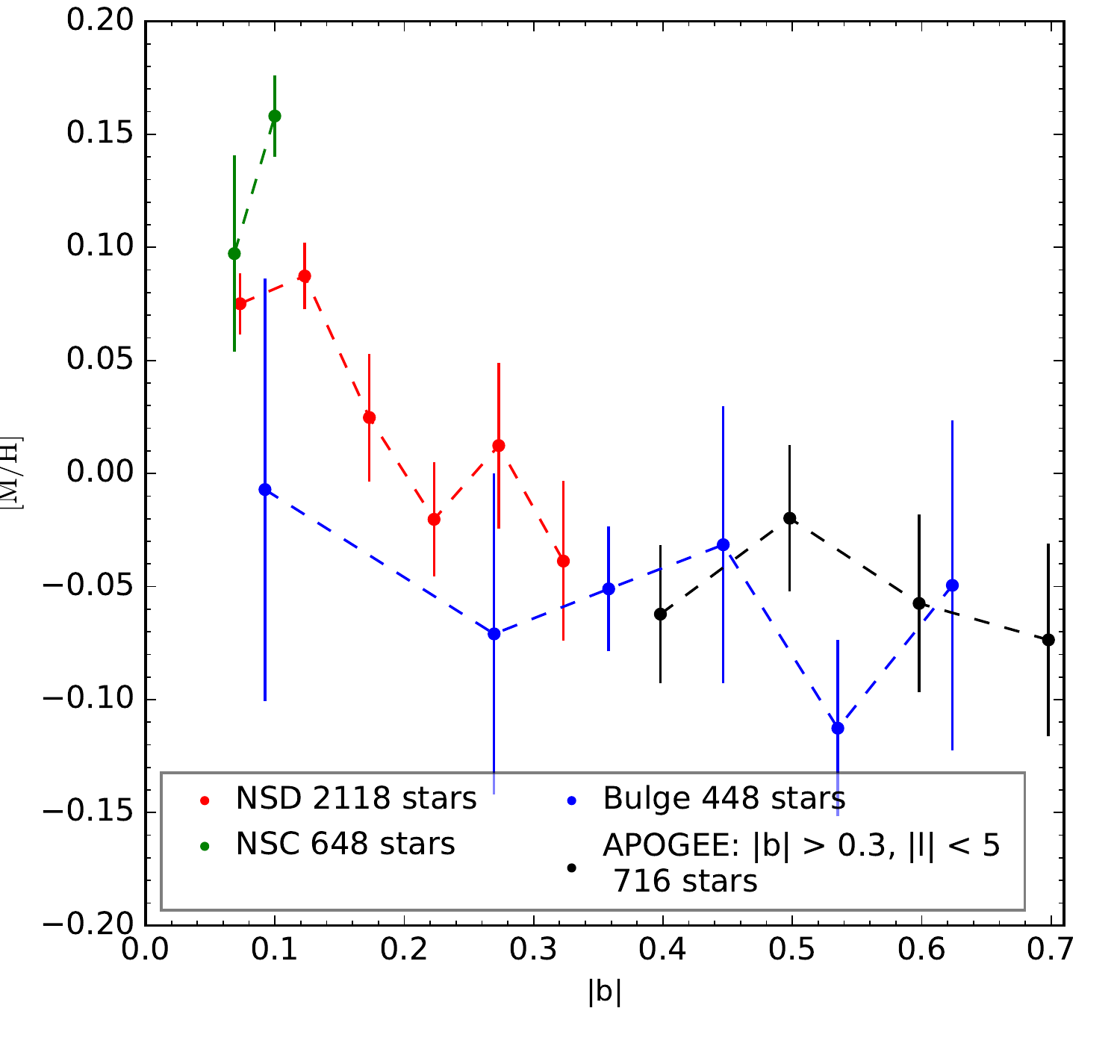}
          \caption{Mean metallicity vs. absolute Galactic latitude. The red points denote the NSD, blue points our bulge sample, and in green the NSC sample of FK20. The black points show the APOGEE DR16 bulge sample.} 
        \label{gradient}
\end{figure}

In Fig.~\ref{gradient} we trace the vertical metallicity gradient of each 
of the three components. We firstly note that we see a negative metallicity gradient for the NSD with a slope of $\rm  -0.43 \pm 0.12$\,dex/deg,  
which is much steeper than our comparison  bulge field with a slope of  $\rm  -0.1 \pm 0.08$\,dex/deg.  Owing to the small scale height of the NSD ($\rm \sim 40\,pc$), the vertical gradient could be due to an increasing contamination with bulge stars.
For comparison we also show APOGEE stars in the bulge within $\rm |l| < 5^{o}$, where we see  a slope of $\rm 0.006 \pm  0.09$\,dex/deg,  which is in general in good agreement between APOGEE and our low-resolution bulge sample. As already shown by \citet{Rich2007}, \citet{Paco2018}, and \citet{schultheis2019},  we confirm  the flat metallicity gradient in the 
inner bulge. 
On the contrary, the NSC  shows a distinct behaviour being more metal-rich and no signs of a negative gradient (see also FK20). We also checked for radial (i.e. longitudinal) metallicity gradients by folding the data with respect to the minor axis. We do not detect any gradient and therefore 
the NSD seems to be radially homogeneous in metallicity.

\begin{figure}[!htbp]
  \centering
        \includegraphics[width=0.49\textwidth,angle=0]{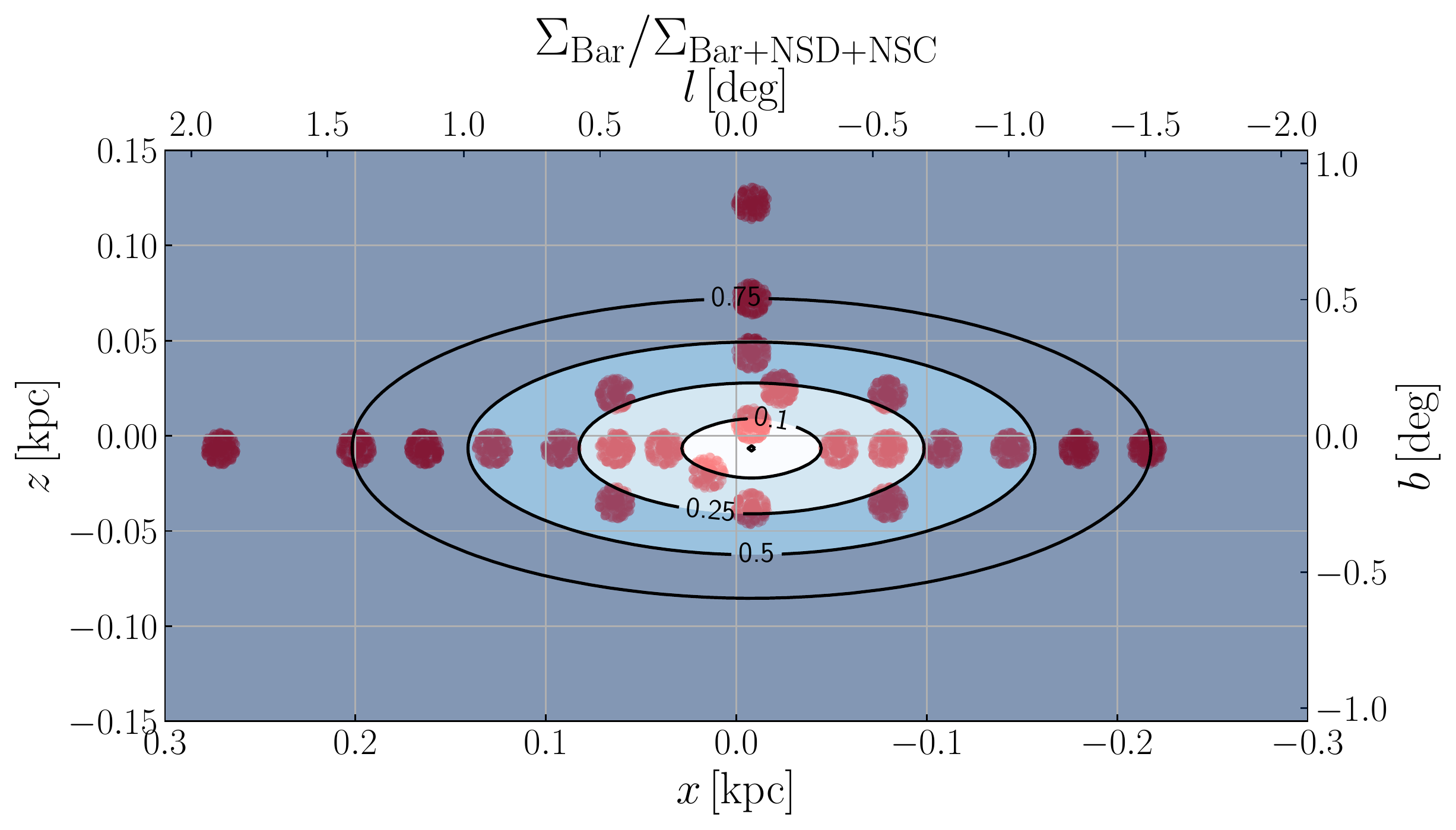}
          \caption{ Surface density ratio between the bulge/bar model from \citet{Launhardt2002} and the sum of the model of the NSC from \citet{Chatzopoulos2015}, the NSD from \citet{Sormani2020} and the bar. This ratio gives a crude estimation of the contamination due to the bar. Our KMOS fields assuming a distance of 8.2\,kpc to the Galactic centre are overplotted.}. 
        \label{contamination}
\end{figure}

To better quantify the contamination of stars not belonging to the NSD (i.e. belonging to the bulge/bar), we use models of the Galactic bulge and bar, the NSD, and the NSC indicated in Fig.~\ref{contamination}. This 
figure shows the surface density ratio of the bulge/bar model from Section 4.2 of \citet{Launhardt2002} to the sum of the fiducial NSD model (model 3) of  \citet{Sormani2020}, the best fitting NSC model of \citet{Chatzopoulos2015}, and the bulge/bar model in \citet{Launhardt2002}.   Assuming that the probability of a given star ending up in our sample is independent of 
the star belonging to the bar/bulge or to the NSD, this ratio can be taken as a crude estimation of the expected contamination due to bulge/bar stars in the sample along each line of sight. While the contamination rate is low in the inner parts of the NSD (ratio smaller than 0.25),  the ratio reaches 0.5 (corresponding to a expected contamination of ~50\%) around longitude ~1 deg, meaning that we expect a significant contamination in the outer parts of the disc.

\section{Metallicity and kinematics}

\begin{figure*}[!htbp]
  \centering
        \includegraphics[width=0.99\textwidth,angle=0]{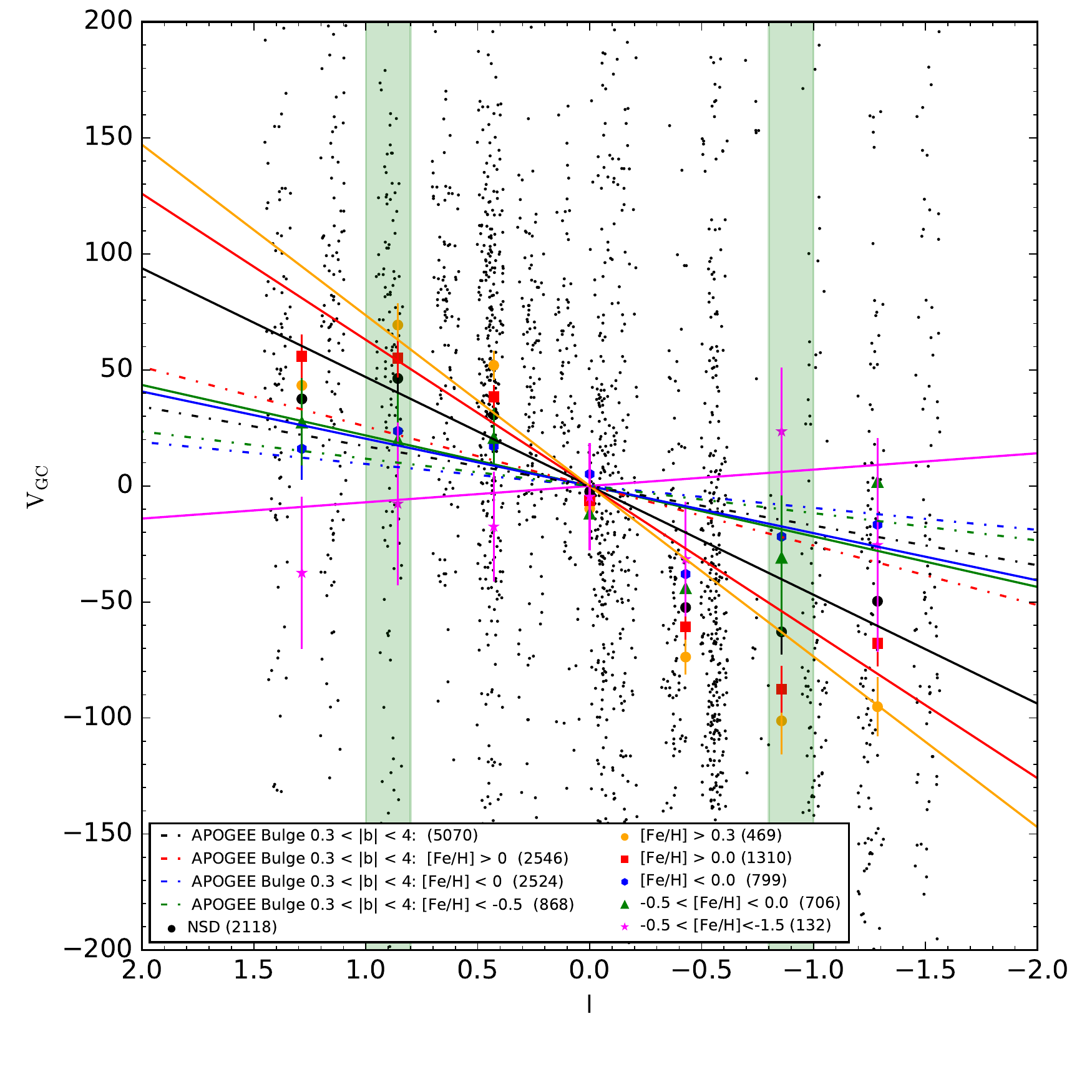}
          \caption{Galactocentric velocity vs.  Galactic longitude.  The points represent the mean values of $\rm V_{GC}$ for samples binned in
Galactic longitude. The respective error bars stand for the standard error of the mean computed as $\sigma/\sqrt{N}$, where N is the number of points in the respective bin. The black line shows a linear fit of the full NSD sample, the orange line  for stars with $\rm [Fe/H] > 0.3$, the red line for 
MR stars with $\rm [Fe/H] > 0$, the blue line for  stars with $\rm [Fe/H] 
< 0$, the green line for stars with   $\rm -0.5 < [Fe/H] < 0$, and the magenta line for stars with   $\rm  [Fe/H] < -0.5$. The dash-dotted lines 
show the corresponding APOGEE bulge sample (see text). The two green areas indicate the approximate radius of the NSD.} 
        \label{rotation}
\end{figure*}

\citet{schoenrich2015} showed based on APOGEE data that the NSD shows a 
clear signature of rotation. Our larger dataset confirms this rotation as 
seen in Fig.~\ref{rotation}. While the black line shows the full dataset, 
we split our sample (see Sect.~3)  into a MR and MP sample. We note at this point a clearly  faster rotation for the MR stars compared to the MP with  the most metal-rich stars in our sample ($\rm [Fe/H] > 0.3\,dex$) showing the 
highest rotation velocities (see Sect.~\ref{CMZ} for a further discussion).
Contrary to \citet{Schultheis2020} and  \citet{schoenrich2015}, we cover in this work the full  spatial range  of  the NSD  allowing  us to trace more precisely the rotation pattern of the NSD. In Tab.~\ref{slope}, we list 
the obtained slopes for different metallicity regimes, their associated errors, and the rotation velocities assuming a disc length of 100\,pc (column 4) and 150\,pc (column 5), respectively.  Assuming a disc length of 100 pc would result in  a rotation velocity of 85\,km/s for the MR part and 30\,km/s for the MP component much slower than found in \citet{Schultheis2020}. Interestingly, stars with $\rm [Fe/H] < -0.5\,dex$ (magenta line) show a very flat rotation with signs of a counter-rotating 
pattern, which could be a sign of a different origin of this metal-poor population such as by accretion from disrupted stellar clusters (\citealt{Tsatsi2017}). The typical uncertainties in the rotational velocities (see 
also column 3 of Tab.~\ref{slope}) is of the order of $\rm \sim 7-10 km/s$.
On the other hand, some of these stars could be  on elongated x1-like orbits that belong to the Galactic bar (\citealt{Molloy2015}, \citealt{Aumer2015}, \citealt{Sormani2015}). We evaluate the statistical significance of the potential negative slope of the velocity trend of these metal-poor stars. We computed 1000 MC resamplings of the points, given their uncertainties, and from them estimated the standard error of the slope. 
Based on the slope
and its standard error, we computed a one-sided $p-$value for the null hypothesis of a flat or positive slope with respect to the alternative
hypothesis of a negative one. The resulting $p-$value of 0.13 implies that we cannot reject the null hypothesis to the 5\% nor to the 10\% of
significance level. To further quantify this, we computed the 70\% confidence interval of the measured slope and found a range of (-13.8,
-0.3), which is consistent with a negative slope to that confidence level. We can see that although our data suggest some evidence of a negative 
slope,
the result is not conclusive and further data will be needed to settle this issue. 
\citet{Do2020} recently found evidence of a kinematically and chemically distinct metal-poor component in the NSC, where the MP component rotates faster than the MR component, which could be a remnant of a massive star cluster formed a few kiloparsec away from the Galactic centre (\citealt{Tsatsi2017}, \citealt{Arca-Sedda2020}). On the other hand, the  metal-poor population could be similar to the non-rotating, old, metal-poor component of the bulge (\citealt{Kunder2020}).

Counter-rotating nuclear stellar discs are also observed in extragalactic 
systems such as the central disc of NGC 4458, which is counter-rotating. This disc is old, metal-rich, and alpha-enhanced (\citealt{Morelli2004}). The NSD of  NGC 4478 is also counter-rotating with respect to its host galaxy but is younger and more metal-rich. To account for the counter-rotation, an external origin of the gas has been  claimed. (\citealt{Morelli2004}). Precise proper motion measurements together with detailed dynamical modelling
and resulting orbital calculations  would be essential to trace the origin
of these metal-poor stars. 

\begin{table}[!htbp]
\begin{tabular}{cccccc}
$\rm [Fe/H]$& slope&$\sigma_{slope}$&$\rm v_{Rot}^{R=100pc}$&$\rm v_{Rot}^{R=150pc}$\\
\hline
full&46.9&8.7&67.7&45.1\\
$\rm -0.5 < [Fe/H] < 0 $&20.2&9.7&29.1&19.4\\
$\rm [Fe/H] < 0$&20.3&6.7&29.6&19.5\\
$\rm [Fe/H] > 0$&62.9&9.8&90.8&60.5\\
$\rm [Fe/H]>0.3$& 73.5&13.2&106.1&70.7\\ 
$\rm -1.5< [Fe/H] < -0.5$&-7.0&10.5&-10.1&-6.7\\
\end{tabular}
\caption{Slope vgc/l  and errors for different metallicities (see  Fig. 8) in $\rm  km\,s^{-1}/deg $. The fourth and fifth columns give the rotational velocities (km/s) assuming a disc length of 100 pc and 150 pc, respectively. }
\label{slope}
\end{table}

For comparison, we show in Fig.~\ref{rotation} a Galactic bulge sample of 
APOGEE DR16 stars selected in the same way as described in \citet{alvaro2020}, covering  the latitude range  $\rm 0.3^{o} < |b| < 4^{o}$ and $\rm |l| < 5^{o}$ and cutting at the galactocentric distance of $\rm R_{GC} < 3.5\,kpc $; this ensures a pure inner  bulge sample in which we exclude the NSD component.  The APOGEE bulge sample shows a small difference in the rotation velocity between metal-rich and metal-poor stars.  We find a slope of $\rm 25.6 \pm 3$ $\rm  km\,s^{-1}/deg $  for the metal-rich stars, while for stars with $\rm [Fe/H] < 0$ and $\rm [Fe/H] < -0.5$ a slope of $\rm 11.7 \pm 3.0 $ and $\rm 9.5 \pm 5.0$  $\rm  km\,s^{-1}/deg $, respectively.
Recent studies of the rotation curve of the Galactic bulge already revealed this  difference in the rotation velocity as a function of metallicity, that is metal-rich stars rotate faster than metal-poor stars (see e.g. \citealt{Clarkson2018}, \citealt{alvaro2020}). \citet{Clarkson2018} speculate  that this could be due  to the bar formation,  which would lead to a higher fraction of metal-rich stars showing more elongated orbits compared to metal-poor stars.  This would be also consistent with the fact that the X-shape structure of the bulge contains mostly metal-rich stars, indicating different orbital motions between MR and MP stars (see e.g. \citealt{babusiaux2010}, \citealt{Ness2013}, \citealt{Alvaro2017}).  However, based on Fig.~\ref{rotation} we  see a much more  pronounced difference in the rotation velocity between the MR and the MP population indicating   that the NSD  shows a  kinematically   distinct structure with respect to the  Galactic bulge with a much faster rotation than the bulge. We also  see  a clear drop in the galactocentric velocities at around $\rm  |l|=1^{o}$   (shown in Fig.~\ref{rotation} as the green area),  which indicates 
 the  radius  of the NSD.    We see a similar feature from star counts ( \citealt{nishiyama13})   declining steeply from 0.6--0.8 degrees  outwards, suggesting a smaller disc length.  The MP population seems to be dominated by  bulge-like kinematics (see Fig.~\ref{rotation}, Fig.~\ref{dispFeH}), which is very similar to that seen  in Galactic bulge studies (e.g. ~\citealt{alvaro2020}). In order to disentangle the combined MP population of bulge and NSD, proper motions would be necessary.
 


\begin{figure}[!htbp]
  \centering
        \includegraphics[width=0.49\textwidth,angle=0]{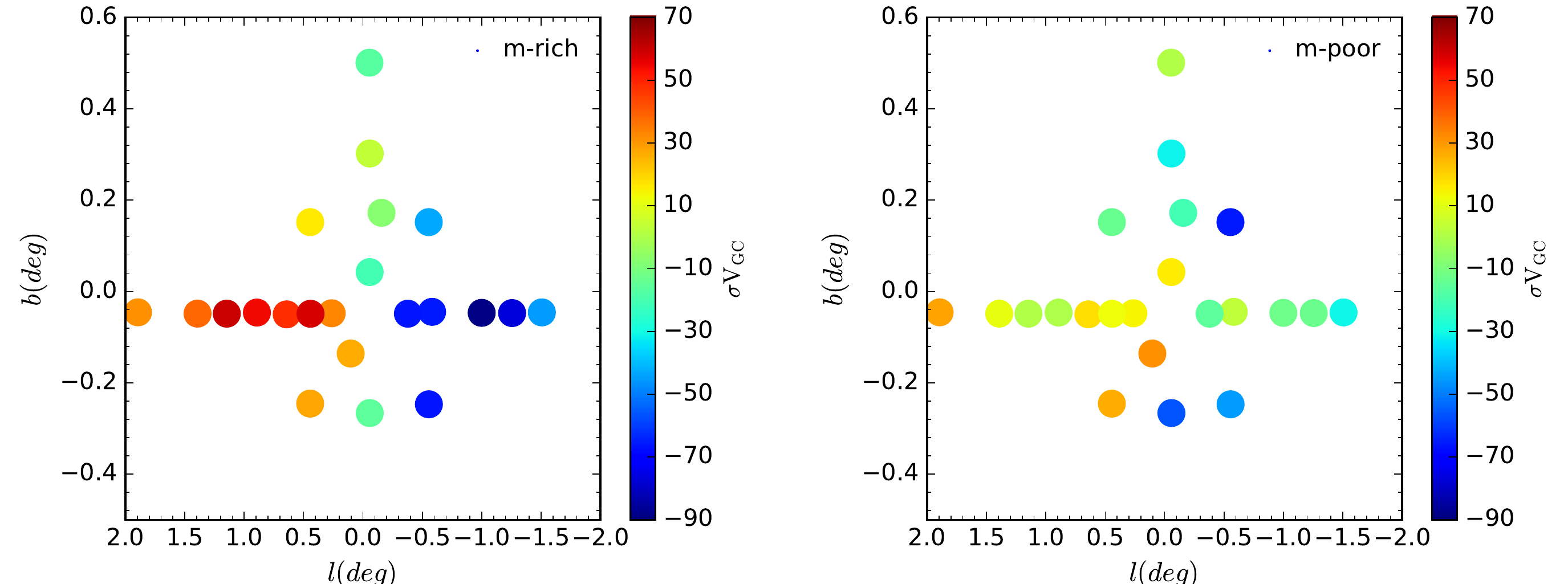}
         \includegraphics[width=0.49\textwidth,angle=0]{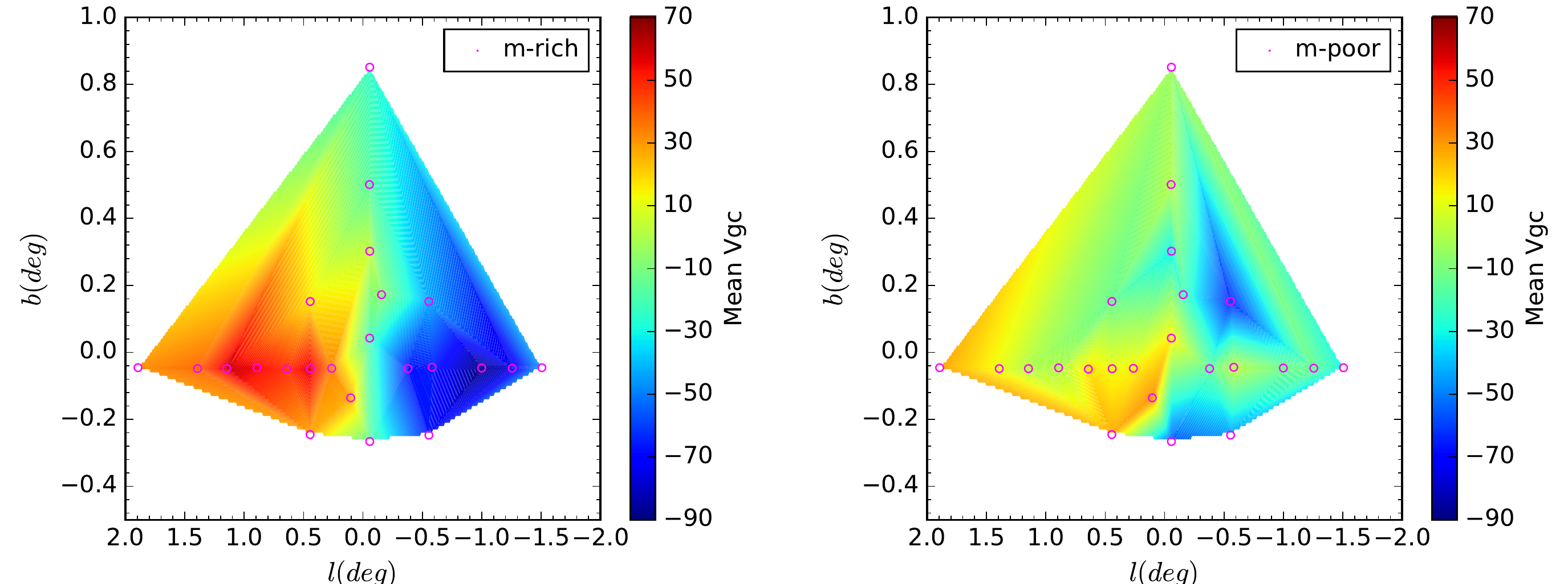}
        \caption{(Upper panels): Mean galactocentric velocity for  MR (left panel) and MP stars (right panel) for each of our KMOS fields. (Lower panels): Similar map but using a linear interpolation to  a rectangular grid. } 
        \label{meanvgc}
\end{figure}

Figure~\ref{meanvgc}  shows the galactocentric radial velocity  $\rm v_{GC}$ of our sample separated into a MR sample (left panel) and a MP sample 
(right panel). For illustration purposes we also show a linear interpolation of our dataset using a rectangular grid and displaying the contour surfaces.  The MR sample shows  a clearly  pronounced rotation pattern around the minor axis. This pattern is a well-known characteristic for the Galactic bulge studies (see e.g. \citealt{Zoccali2014}, \citealt{Zasowski2016}) and for the latest updated APOGEE data release by \citet{alvaro2020}. However, we note a much more  uniform rotation pattern with a much slower rotation  for the MP component, as shown in Fig.~\ref{rotation}.

\begin{figure}[!htbp]
  \centering
        \includegraphics[width=0.49\textwidth,angle=0]{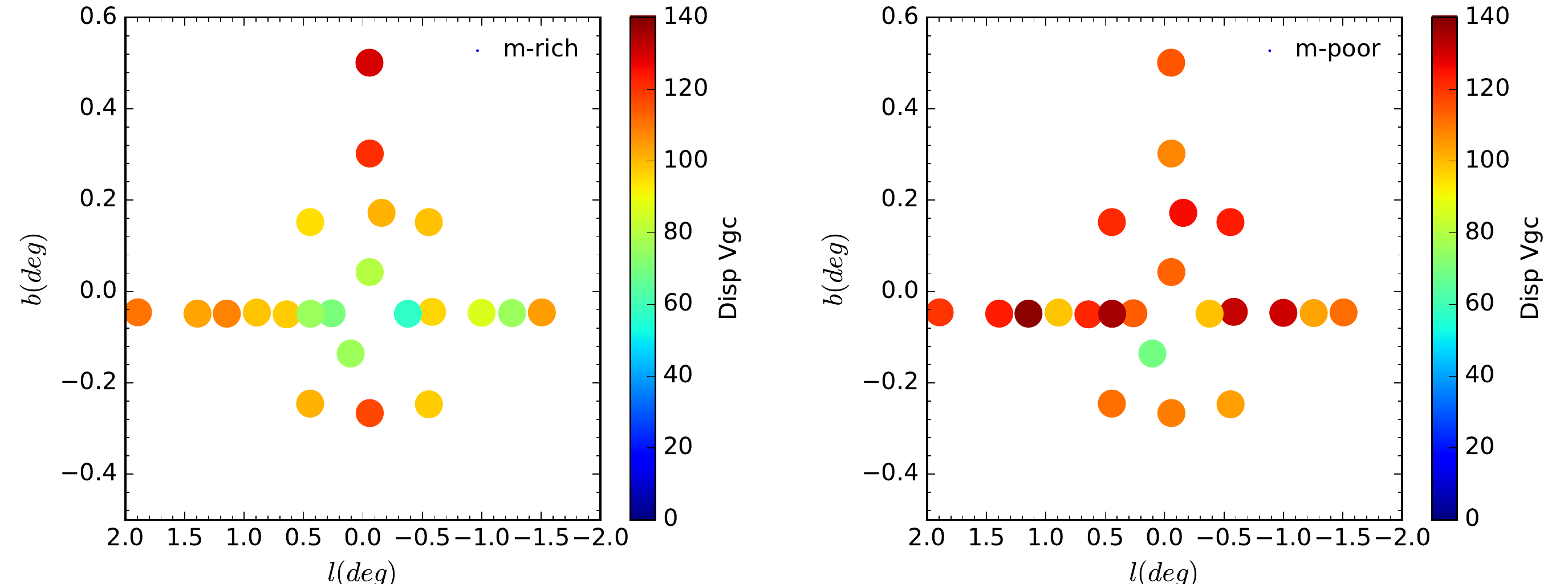}
        \includegraphics[width=0.49\textwidth,angle=0]{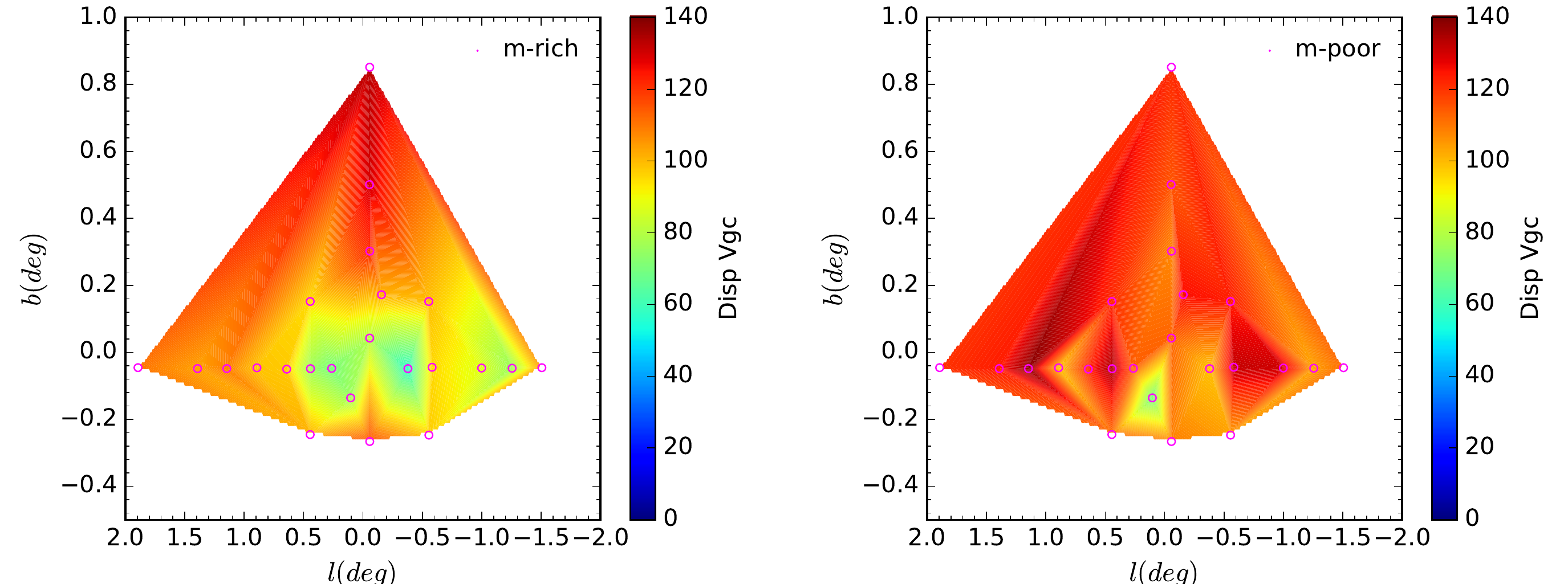}
        \caption{Galactocentric velocity dispersion for  MR (left panel) and MP stars (right panel) } 
        \label{dispvgc}
\end{figure}

The patterns of the velocity dispersion (see Fig.~\ref{dispvgc}) also show some significant differences between the MR and MP components. In general, we see that the velocity dispersion decreases with increasing metallicity, that is the most metal-rich stars are dynamically cool.  A similar picture on a more global scale extended to the Galactic bulge was carried out by \citet{zoccali2017} using GIBS data,  whereas in this work we  concentrate 
on the inner degree where the GIBS data  only have one data point. Our map (see lower panel of Fig.~\ref{dispvgc})  nicely shows some  structures with low velocity dispersion for the MR population  within $\rm 0.1^{o}$ in Galactic latitude  where the dispersion is highest at higher Galactic latitudes with a pronounced enhancement at around $\rm b > 0.4^{o}$. This would be the spatial limit of the NSD, indicating that the NSD might have in general a lower velocity dispersion and is therefore kinematical cooler 
than the Galactic bulge. The MP population shows in general a  higher radial velocity dispersion  in the bulge, but shows for the NSD some low velocity dispersion  windows that match the kinematically cool component of the MR part in the NSD.

 \begin{figure}[!htbp]
  \centering
        \includegraphics[width=0.49\textwidth,angle=0]{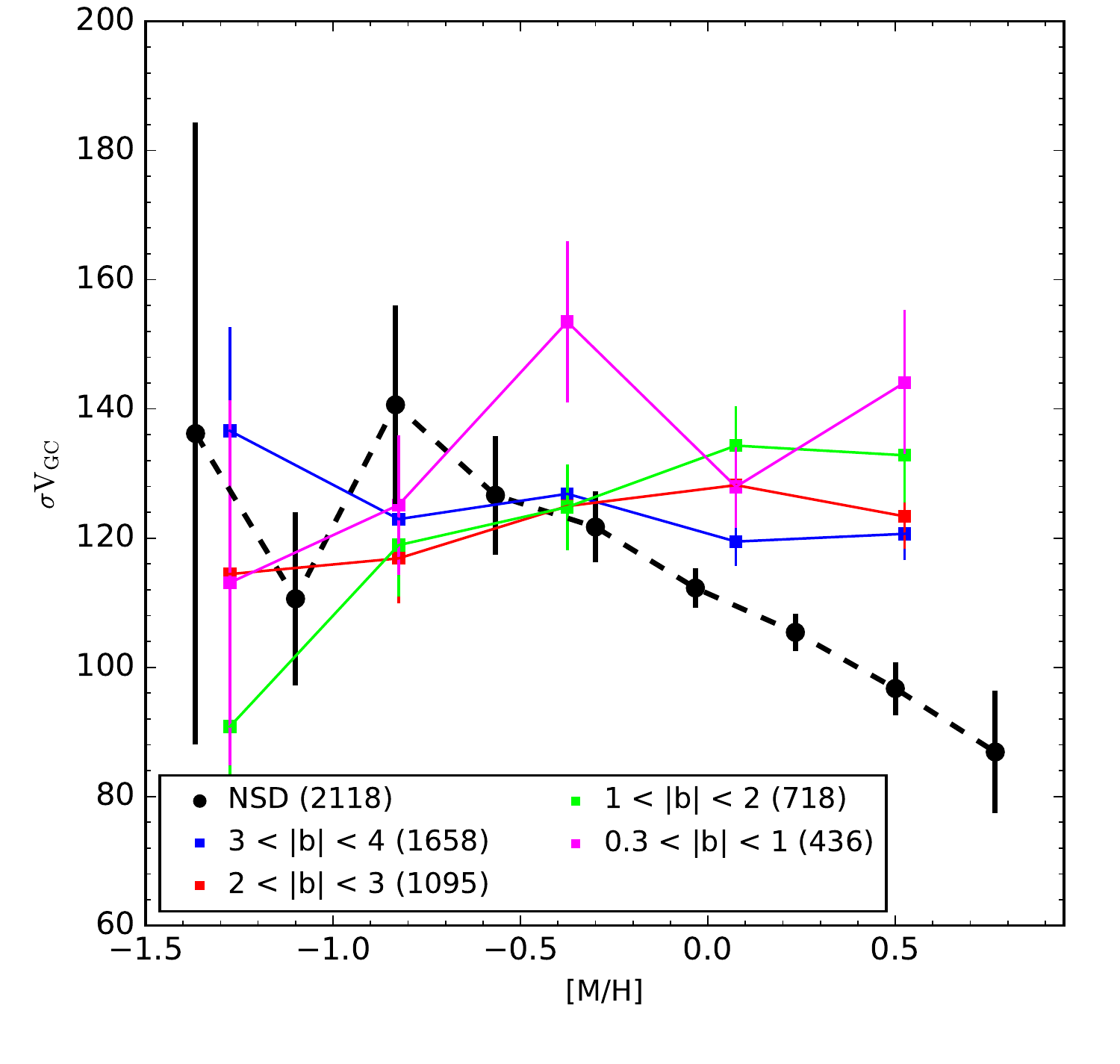}
        \caption{Galactocentric velocity dispersion as a function of metallicity. Filled lines denote APOGEE DR16 bulge data: $\rm 0.3 < |b|< 1^{o}$ (magenta),  $\rm 1 < |b|< 2^{o}$ (green), $\rm 2 < |b|< 3^{o}$ (red), and $\rm 3 < |b| < 4^{o}$ (blue). The dashed black line shows our NSD sample} 
        \label{dispFeH}
\end{figure}

\begin{figure*}[!htbp]
  \centering
  \includegraphics[width=0.49\textwidth,angle=0]{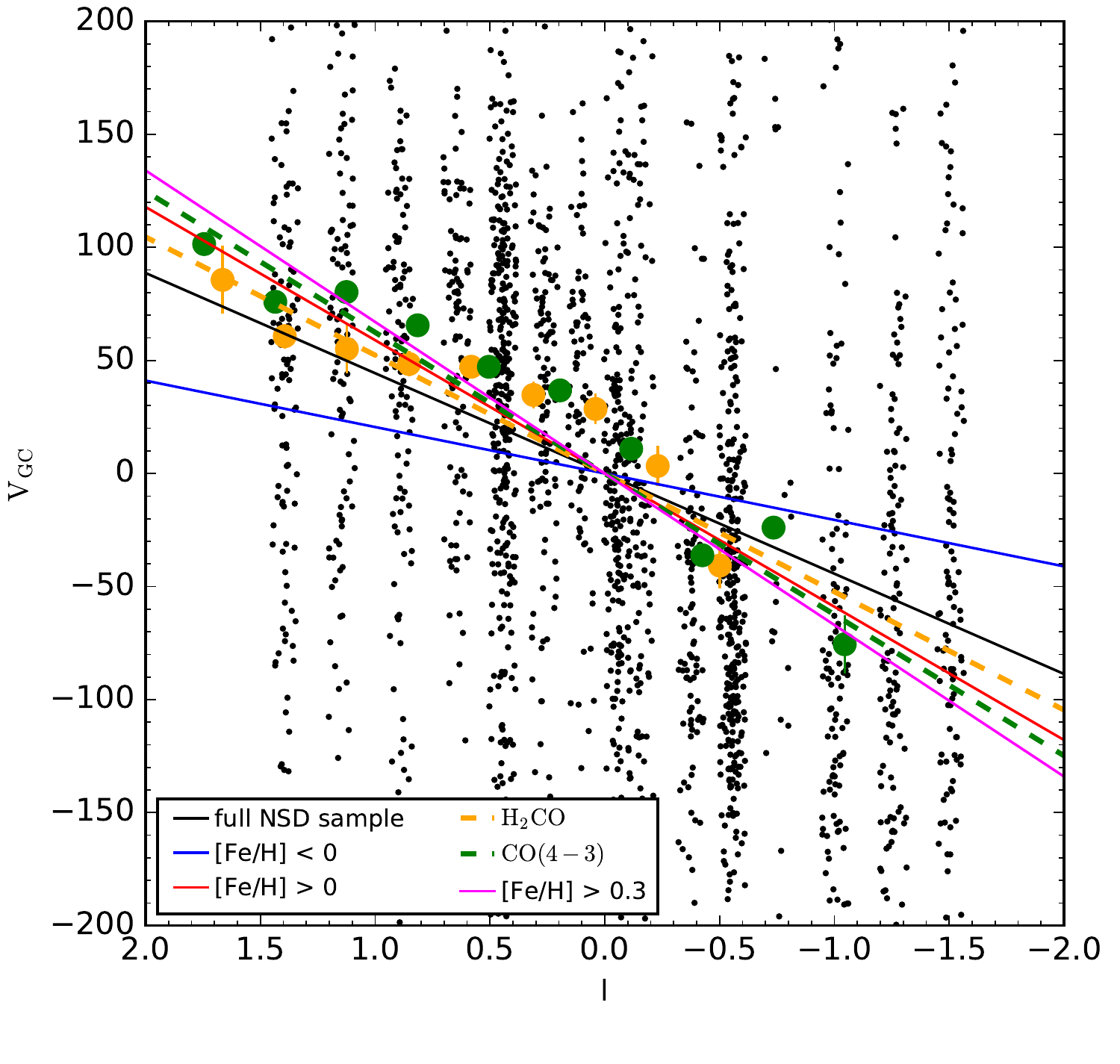}
  \includegraphics[width=0.49\textwidth,angle=0]{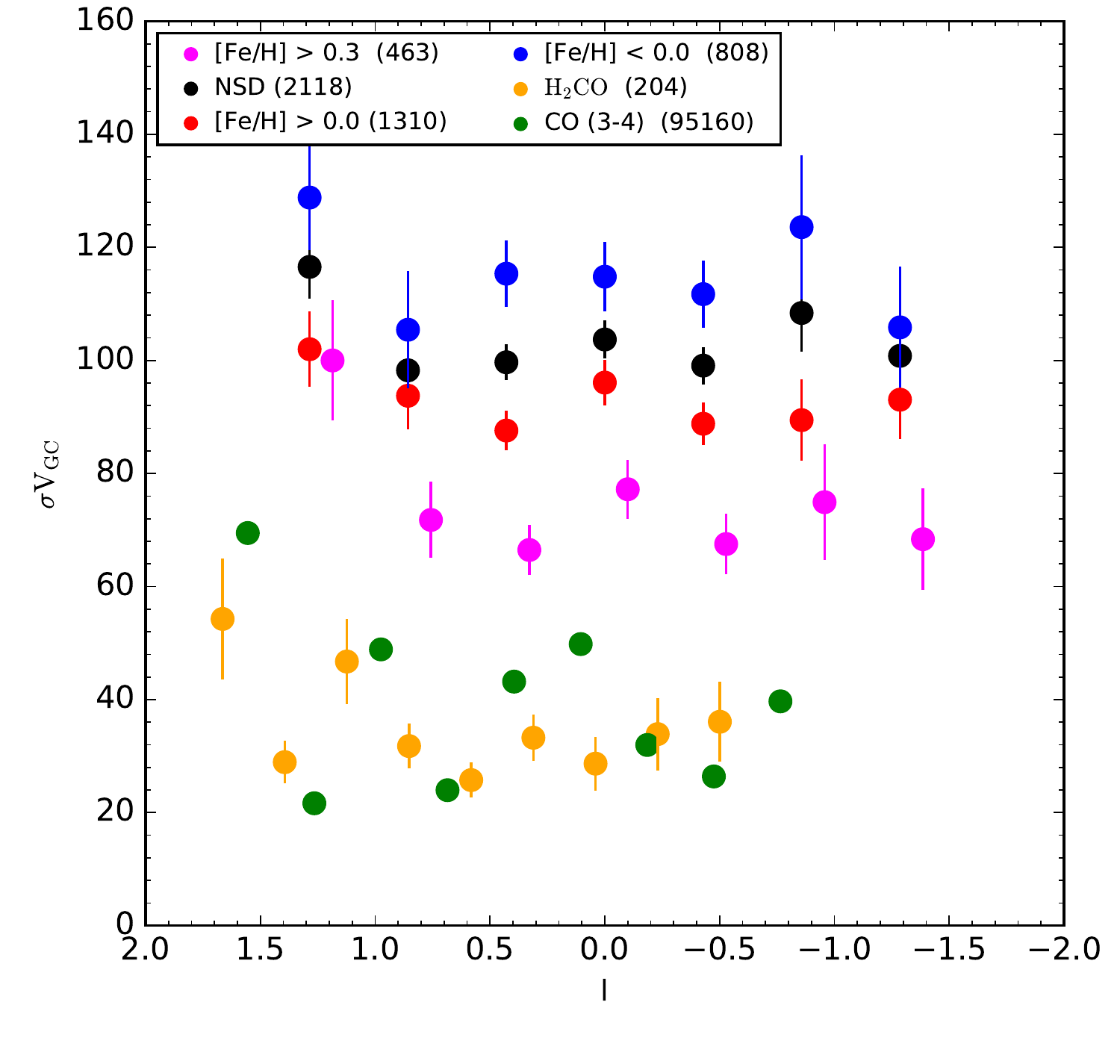}
        \caption{(Left panel): Similar as in Fig.~8. The magenta line indicates the fitted relation for $\rm [Fe/H] > 0.3$. The orange points show the $\rm H_{2}CO$ data of \protect  \citet{Ginsburg2016} and the orange dotted line shows the best linear fit. (Right panel): Velocity dispersion as a function of Galactic longitude (same symbols).}
        \label{rotgaz}
\end{figure*}

We then ask if we see any trend of the kinematic behaviour with the global metallicity. We trace in Fig.~\ref{dispFeH} the velocity dispersion as a function of  metallicity. As a comparison sample we take the latest APOGEE DR16 dataset (\citealt{Ahumada2019}) and constructed our bulge sample by applying 
a cut in  galactocentric distance $\rm R_{GC} < 3.5\,kpc$  in the same way as described in \citet{alvaro2020}.  We show in Fig.~\ref{dispFeH} the APOGEE bulge data sample at different  Galactic latitudes, $\rm 0.3< |b|< 
1^{o}$,  $\rm 1 < |b|< 2^{o}$, $\rm 2 < |b|< 3^{o}$, and $\rm 3 < |b| < 4^{o}$,  and our NSD sample in black. As already noted by \citet{zoccali2017} and \citet{alvaro2020}, we see a higher velocity dispersion for metal-poor stars than for metal-rich stars for stars at $\rm |b| >3-4^{o}$. For $\rm 2< |b|< 3^{o}$  we see that the curves get flatter resulting in a similar velocity dispersion for MP and MR stars, while for  $\rm 0.3< |b|< 2 ^{o}$   we clearly see  an inverse trend in the sense that the velocity dispersion increases with  higher metallicity.  A similar trend is already noted by \citet{zoccali2017}  where they find based on GIBS data a reverted behaviour for stars closer 
to the plane at around $\rm b = -1^{o}$.  Their sample is just split up into two metallicity bins (MR and MP with super and sub-solar metallicities). However Fig.~\ref{dispFeH} shows  the trend in absolute metallicities ,
where  our NSD sample is indicated by the black-dashed  line  showing an opposite trend in the sense that the velocity dispersion decreases with increasing  metallicity and where clearly the bulk of our metal-rich stars are kinematically cool compared to the bulge sample. This is a strong indication that the NSD is chemically and kinematically 
a distinct component from the Galactic bulge for  metallicities $\rm [M/H] > -0.5$, while for metal-poorer stars with $\rm [M/H] < -0.5$ the correlation between metallicity and velocity dispersion overlaps with the Galactic bulge.

\subsection{Possible formation of metal-rich stars of the NSD in the CMZ} \label{CMZ}

The CMZ  is spatially associated with the NSD. The molecular gas in the CMZ is characterised by its  high kinetic temperature and a  much higher density compared to other star forming regions in 
the Galaxy. To compare the kinematics of the gas with that of the stellar content, we use two tracers of the gas: First, the $\rm H_{2}CO$ molecule, which is ubiquitous and traces all the molecular gas, especially 
in high density regions such as the CMZ (\citealt{AO2013}).  \citet{Ginsburg2016} mapped the inner 300\,pc of the CMZ with the J=3--2 para-formaldehyde (p-$\rm H_{2}CO$) transition using the APEX-1 instrument on the Atacama Pathfinder Experiment (APEX) telescope. These data are publicly available and we retrieved the positions (in Galactic coordinates)  and the  velocities, which have been transformed to the galactocentric velocities. The spectral resolution is about  30\,$\arcsec$  with about 1 km/s resolution in velocity. Second, the CO (J=4--3) data at 461.041 GHz from  the AST/RO survey of  \citet{Martin2004}, which is found to be coextensive with lower J transitions of CO (e.g J=1-0).  The data span the longitude range of $\rm =-1.3^{o} < l < 2^{o}$ and a latitude range of $\rm -0^{o}.3 < b < 0^{o}.2$.
The angular  resolution is about 2\arcmin, which is lower than other general surveys in the GC region; but these data have the advantage that they are  publicly available. We used the  FITS datacubes and extracted along each pixel in (l,b) the velocity information, which  we again transformed to the galactocentric system.


Figure~\ref{rotgaz} shows  a similar diagram as in Fig.~\ref{rotation}, where we now include the molecular gas of the CMZ,
both the  CO (J=4-3)  and the $\rm H_{2}CO$ molecule.  
We see that both tracers show  very similar rotation pattern (left panel) 
with similar velocity dispersions (right panel), although the CO molecule shows a larger variation in the velocity dispersion compared to $\rm H_{2}CO,$ which  could be due to  the lower spatial resolution  of the CO(J=4-3) data. In addition, we show the rotation curve of super metal-rich stars ($\rm [Fe/H] >0.3\dex$) in magenta.
By comparing the rotation velocities of the molecular gas, we  see from Fig.~\ref{rotgaz} (left panel) that the molecular gas  clearly shows similar rotation patterns to our MR populations.  This could be an indication that the MR stars originate from the CMZ.  On the right panel the velocity dispersions of  our NSD sample as well of the molecular gas are shown where we see that the mean velocity dispersion of the gas is of the 
order of 30-40 km/s, which is  in very good in agreement with other literature studies (see e.g. Morris\& Serabyn 1996, G\"usten \& Philipp 2004), while
the MR stars are kinematically hotter with velocity dispersions of the order of $\rm \sim 70-90\,km/s$.
\citet{Ted2019} trace age-velocity dispersion relation in the disc and find that for the inner disc the stellar populations are dynamically heated both by giant molecular clouds and spiral arms. The CMZ is well known to have a high density of molecular clouds with unusually high densities on spatial scales of about 1\,pc (\citealt{Kauffmann2017}). This could be 
the cause of the higher dispersion of the stars compared to the gas.

\citet{nogueras2020} find some evidence of a recent star formation burst 
1 Gyr ago, which could be linked to  the formation of the most metal-rich stars. Measurements of the interstellar medium in the Galactic centre also indicate super -solar metallicities (\citealt{Shields1994}, \citealt{Maeda2002}).


\subsection{Formation  mechanisms and their link to extragalactic nuclear discs}

Galactic bars play a major role in galaxy formation and evolution. They have been identified at high redshifts (\citealt{simmons2014}),  suggesting that bars influence the evolution of  their host galaxies over long timescales up to 10\,Gyr. There is now more and more proof that the bar drives the formation of the so-called boxy/peanut bulges  owing to the secular internal evolution of the disc  (\citealt{Combes1990}). In addition, bars 
can create substructures in the nuclear region of disc galaxies such as nuclear discs and nuclear rings by redistributing angular momentum (\citealt{Combes1985}). There is now growing evidence that nuclear  discs and nuclear rings are built from gas that was funnelled to the centre by the bar (see e.g. \citealt{Athanassoula1992}, \citealt{Kim2011}, \citealt{Fragkoudi2016}). However, it is still unclear which physical mechanism determines the size of these structures. \citet{Gadotti2020} find correlations between the bar length and the size of nuclear discs and nuclear rings.

Another formation scenario of nuclear discs could be initiated by galaxy mergers (\citealt{Mayer2008}, \citealt{Eliche-Moral2011} \citealt{Chapon2013}). However, galaxy mergers fail to reproduce the sizes of typically observed nuclear discs in nearby galaxies (such as the MUSE TIMER project \citealt{Gadotti2019}) and  they cannot reproduce the detailed kinematical maps as 
 discussed in \citet{Gadotti2020}.

In our work, we see that metal-poor stars show a very similar kinematical behaviour with as the Galactic bulge and it is most likely that the metal-poor population of the NSD overlaps strongly with the Galactic bulge. The metal-rich stars
($\rm [Fe/H] > 0$)  are dynamically cold  with a must faster rotation pattern than the bulge and they seem to show a very distinct kinematics compared to the bulge population. This is what is expected if the nuclear disc has formed  from a bar-driven secular evolution process.

We see a similar behaviour when comparing with extragalactic nuclear discs.  \citet{Gadotti2020} find that extragalactic nuclear discs  have dynamically distinct features with respect to the main galaxy disc:  faster rotation and lower velocity dispersion with near-circular orbits. 
The TIMER survey  also indicates  that the nuclear disc is a separate component from the main galactic disc. In terms of chemistry, the TIMER project suggests (\citealt{Bittner2020}) that nuclear discs are metal-richer and 
younger with  lower $\rm [\alpha/Fe]$ abundances than their surroundings (see their Fig.~5).

Age dating such as isotopic ratios of $\rm C^{12}/C^{13}$ as well as  $\rm [\alpha/Fe]$ elements and iron-peak elements   using high-resolution spectroscopy would be essential to study the detailed formation history of the 
NSD. The forthcoming new generation of instruments such as CRIRES+ or MOONS at the VLT will allow us to determine ages and therefore trace the detailed SFH in the NSD.


\section{Conclusions}
Using  a KMOS survey of 2118 stars located in the NSD, We trace the metallicities and kinematics by comparing this survey with the NSC and a bulge comparison sample. We  find that
\begin{enumerate}[i)]
  
\item  The MDF of the NSD is more metal-rich than the bulge sample and less metal-rich than the NSC sample. A KS test confirms that these three samples do not come from the same population.
\item  Splitting the sample into  metal-rich ($\rm [Fe/H] > 0$) and metal-poor populations ($\rm [Fe/H] < 0$), the fraction of MR to MP stars is about 1 for the bulge, while this fraction increases to 1.6 for the NSD and reaches a 
peak of 2.6 for the NSC. This confirms the distinction in chemistry for these 
three populations.
\item  Bulge stars  are in general  kinematically hot while the NSD shows 
a kinematically cool component, where the velocity dispersion decreases with increasing metallicity contrary to the Galactic bulge.
\item   The metal-rich population rotates faster than the metal-poor component.
  \end{enumerate}
We use  molecular gas tracers such as CO(3-4) and $\rm H_{2}CO$  and find 
evidence  that the gas rotation velocity corresponds suprisingly well with  those of the  MR population in the NSD. This is a strong argument that 
the metal-rich stars  have been formed  in the CMZ, while in contrast the metal-poor population with a much lower rotation velocity might have a different origin.
We conclude that the NSD is chemically and kinematically distinct to the Galactic bulge  and the NSC, and that its formation is most likely not related to the NSC or the bulge. Comparing our results with the TIMER survey, we find similar properties of extragalactic nuclear discs and the Milky Way  NSD, that is a  dynamically cold, fast rotating and metal-rich distinct  component from the surrounding disc or bulge.

This agrees with the findings of \citet{nogueras2020}, in which they 
clearly identify a different and very specific SFH compared to the Galactic bulge and the NSC. In the  future, precise chemical abundances in the NSD will be required to confirm our findings.

\begin{acknowledgements}
  MS  acknowledges the Programme National de Cosmologie et Galaxies (PNCG) of CNRS/INSU, France, for financial support.  We want to thank Karl Menten for discussing the molecular gas data. R. S.  acknowledges financial support from the State Agency for Research of the Spanish MCIU through the “Center of Excellence Severo Ochoa” award for the Instituto de Astrofísica de Andalucía (SEV-2017-0709).  R. S. acknowledges financial support from national project PGC2018-095049-B-C21 (MCIU/AEI/FEDER, UE). FNL and NN acknowledge funding by the Deutsche Forschungsgemeinschaft (DFG, German Research Foundation) -- Project-ID 138713538 -- SFB 881 (``The Milky Way System'', subproject B08). MCS acknowledges financial support from the German Research Foundation (DFG) via the collaborative research center (SFB 881,Project-ID 138713538) “The Milky Way System” (subprojects B1, B2, and B8) and from the European Research Council via the ERC Synergy Grant “ECOGAL – Understanding our Galactic ecosystem: from the disc of the Milky Way to the formation sites of stars and planets” (grant855130). ARA acknowledges support from FONDECYT through grant 3180203.

\end{acknowledgements}


\bibliographystyle{aa}
\bibliography{nsd_accepted}

\begin{thebibliography}{77}
\expandafter\ifx\csname natexlab\endcsname\relax\def\natexlab#1{#1}\fi

\bibitem[{{Ahumada} {et~al.}(2020){Ahumada}, {Allende Prieto}, {Almeida},
  {Anders}, {Anderson}, {Andrews}, {Anguiano}, {Arcodia}, {Armengaud},
  {Aubert}, {Avila}, {Avila-Reese}, {Badenes}, {Balland}, {Barger},
  {Barrera-Ballesteros}, {Basu}, {Bautista}, {Beaton}, {Beers}, {Benavides},
  {Bender}, {Bernardi}, {Bershady}, {Beutler}, {Bidin}, {Bird}, {Bizyaev},
  {Blanc}, {Blanton}, {Boquien}, {Borissova}, {Bovy}, {Brandt}, {Brinkmann},
  {Brownstein}, {Bundy}, {Bureau}, {Burgasser}, {Burtin}, {Cano-D{\'\i}az},
  {Capasso}, {Cappellari}, {Carrera}, {Chabanier}, {Chaplin}, {Chapman},
  {Cherinka}, {Chiappini}, {Doohyun Choi}, {Chojnowski}, {Chung}, {Clerc},
  {Coffey}, {Comerford}, {Comparat}, {da Costa}, {Cousinou}, {Covey}, {Crane},
  {Cunha}, {da Silva Ilha}, {Dai}, {Damsted}, {Darling}, {Davidson}, {Davies},
  {Dawson}, {De}, {de la Macorra}, {De Lee}, {de Andrade Queiroz}, {Deconto
  Machado}, {de la Torre}, {Dell'Agli}, {du Mas des Bourboux},
  {Diamond-Stanic}, {Dillon}, {Donor}, {Drory}, {Duckworth}, {Dwelly},
  {Ebelke}, {Eftekharzadeh}, {Eigenbrot}, {Elsworth}, {Eracleous},
  {Erfanianfar}, {Escoffier}, {Fan}, {Farr}, {Fern{\'a}ndez-Trincado},
  {Feuillet}, {Finoguenov}, {Fofie}, {Fraser-McKelvie}, {Frinchaboy},
  {Fromenteau}, {Fu}, {Galbany}, {Garcia}, {Garc{\'\i}a-Hern{\'a}ndez}, {Garma
  Oehmichen}, {Ge}, {Geimba Maia}, {Geisler}, {Gelfand}, {Goddy},
  {Gonzalez-Perez}, {Grabowski}, {Green}, {Grier}, {Guo}, {Guy}, {Harding},
  {Hasselquist}, {Hawken}, {Hayes}, {Hearty}, {Hekker}, {Hogg}, {Holtzman},
  {Horta}, {Hou}, {Hsieh}, {Huber}, {Hunt}, {Ider Chitham}, {Imig}, {Jaber},
  {Jimenez Angel}, {Johnson}, {Jones}, {J{\"o}nsson}, {Jullo}, {Kim},
  {Kinemuchi}, {Kirkpatrick}, {Kite}, {Klaene}, {Kneib}, {Kollmeier}, {Kong},
  {Kounkel}, {Krishnarao}, {Lacerna}, {Lan}, {Lane}, {Law}, {Le Goff}, {Leung},
  {Lewis}, {Li}, {Lian}, {Lin}, {Long}, {Longa-Pe{\~n}a}, {Lundgren}, {Lyke},
  {Ted Mackereth}, {MacLeod}, {Majewski}, {Manchado}, {Maraston}, {Martini},
  {Masseron}, {Masters}, {Mathur}, {McDermid}, {Merloni}, {Merrifield},
  {M{\'e}sz{\'a}ros}, {Miglio}, {Minniti}, {Minsley}, {Miyaji}, {Mohammad},
  {Mosser}, {Mueller}, {Muna}, {Mu{\~n}oz-Guti{\'e}rrez}, {Myers}, {Nadathur},
  {Nair}, {Nandra}, {do Nascimento}, {Nevin}, {Newman}, {Nidever}, {Nitschelm},
  {Noterdaeme}, {O'Connell}, {Olmstead}, {Oravetz}, {Oravetz}, {Osorio},
  {Pace}, {Padilla}, {Palanque-Delabrouille}, {Palicio}, {Pan}, {Pan},
  {Parker}, {Paviot}, {Peirani}, {Pe{\~n}a Ram{\'r}ez}, {Penny}, {Percival},
  {Perez-Fournon}, {P{\'e}rez-R{\`a}fols}, {Petitjean}, {Pieri},
  {Pinsonneault}, {Poovelil}, {Povick}, {Prakash}, {Price-Whelan}, {Raddick},
  {Raichoor}, {Ray}, {Rembold}, {Rezaie}, {Riffel}, {Riffel}, {Rix}, {Robin},
  {Roman-Lopes}, {Rom{\'a}n-Z{\'u}{\~n}iga}, {Rose}, {Ross}, {Rossi},
  {Rowlands}, {Rubin}, {Salvato}, {S{\'a}nchez}, {S{\'a}nchez-Menguiano},
  {S{\'a}nchez-Gallego}, {Sayres}, {Schaefer}, {Schiavon}, {Schimoia},
  {Schlafly}, {Schlegel}, {Schneider}, {Schultheis}, {Schwope}, {Seo},
  {Serenelli}, {Shafieloo}, {Shamsi}, {Shao}, {Shen}, {Shetrone}, {Shirley},
  {Silva Aguirre}, {Simon}, {Skrutskie}, {Slosar}, {Smethurst}, {Sobeck},
  {Sodi}, {Souto}, {Stark}, {Stassun}, {Steinmetz}, {Stello}, {Stermer},
  {Storchi-Bergmann}, {Streblyanska}, {Stringfellow}, {Stutz}, {Su{\'a}rez},
  {Sun}, {Taghizadeh-Popp}, {Talbot}, {Tayar}, {Thakar}, {Theriault}, {Thomas},
  {Thomas}, {Tinker}, {Tojeiro}, {Toledo}, {Tremonti}, {Troup}, {Tuttle},
  {Unda-Sanzana}, {Valentini}, {Vargas-Gonz{\'a}lez}, {Vargas-Maga{\~n}a},
  {V{\'a}zquez-Mata}, {Vivek}, {Wake}, {Wang}, {Weaver}, {Weijmans}, {Wild},
  {Wilson}, {Wilson}, {Wolthuis}, {Wood-Vasey}, {Yan}, {Yang}, {Y{\`e}che},
  {Zamora}, {Zarrouk}, {Zasowski}, {Zhang}, {Zhao}, {Zhao}, {Zheng}, {Zheng},
  {Zhu}, \& {Zou}}]{Ahumada2019}
{Ahumada}, R., {Allende Prieto}, C., {Almeida}, A., {et~al.} 2020, \apjs, 249,
  3

\bibitem[{{Ao} {et~al.}(2013){Ao}, {Henkel}, {Menten}, {Requena-Torres},
  {Stanke}, {Mauersberger}, {Aalto}, {M{\"u}hle}, \& {Mangum}}]{AO2013}
{Ao}, Y., {Henkel}, C., {Menten}, K.~M., {et~al.} 2013, \aap, 550, A135

\bibitem[{{Arca Sedda} {et~al.}(2020){Arca Sedda}, {Gualandris}, {Do},
  {Feldmeier-Krause}, {Neumayer}, \& {Erkal}}]{Arca-Sedda2020}
{Arca Sedda}, M., {Gualandris}, A., {Do}, T., {et~al.} 2020, \apjl, 901, L29

\bibitem[{{Athanassoula}(1992)}]{Athanassoula1992}
{Athanassoula}, E. 1992, \mnras, 259, 345

\bibitem[{{Aumer} \& {Sch{\"o}nrich}(2015)}]{Aumer2015}
{Aumer}, M. \& {Sch{\"o}nrich}, R. 2015, \mnras, 454, 3166

\bibitem[{{Baba} \& {Kawata}(2020)}]{Baba2020}
{Baba}, J. \& {Kawata}, D. 2020, \mnras, 492, 4500

\bibitem[{{Babusiaux} {et~al.}(2010){Babusiaux}, {G{\'o}mez}, {Hill}, {Royer},
  {Zoccali}, {Arenou}, {Fux}, {Lecureur}, {Schultheis}, {Barbuy}, {Minniti}, \&
  {Ortolani}}]{babusiaux2010}
{Babusiaux}, C., {G{\'o}mez}, A., {Hill}, V., {et~al.} 2010, \aap, 519, A77

\bibitem[{{Bittner} {et~al.}(2020){Bittner}, {S{\'a}nchez-Bl{\'a}zquez},
  {Gadotti}, {Neumann}, {Fragkoudi}, {Coelho}, {de Lorenzo-C{\'a}ceres},
  {Falc{\'o}n-Barroso}, {Kim}, {Leaman}, {Mart{\'\i}n-Navarro},
  {M{\'e}ndez-Abreu}, {P{\'e}rez}, {Querejeta}, {Seidel}, \& {van de
  Ven}}]{Bittner2020}
{Bittner}, A., {S{\'a}nchez-Bl{\'a}zquez}, P., {Gadotti}, D.~A., {et~al.} 2020,
  \aap, 643, A65

\bibitem[{{Chapon} {et~al.}(2013){Chapon}, {Mayer}, \& {Teyssier}}]{Chapon2013}
{Chapon}, D., {Mayer}, L., \& {Teyssier}, R. 2013, \mnras, 429, 3114

\bibitem[{{Chatzopoulos} {et~al.}(2015){Chatzopoulos}, {Fritz}, {Gerhard},
  {Gillessen}, {Wegg}, {Genzel}, \& {Pfuhl}}]{Chatzopoulos2015}
{Chatzopoulos}, S., {Fritz}, T.~K., {Gerhard}, O., {et~al.} 2015, \mnras, 447,
  948

\bibitem[{{Clarkson} {et~al.}(2018){Clarkson}, {Calamida}, {Sahu}, {Brown},
  {Gennaro}, {Avila}, {Valenti}, {Debattista}, {Rich}, {Minniti}, {Zoccali}, \&
  {Aufdemberge}}]{Clarkson2018}
{Clarkson}, W.~I., {Calamida}, A., {Sahu}, K.~C., {et~al.} 2018, \apj, 858, 46

\bibitem[{{Combes} {et~al.}(1990){Combes}, {Debbasch}, {Friedli}, \&
  {Pfenniger}}]{Combes1990}
{Combes}, F., {Debbasch}, F., {Friedli}, D., \& {Pfenniger}, D. 1990, \aap,
  233, 82

\bibitem[{{Combes} \& {Gerin}(1985)}]{Combes1985}
{Combes}, F. \& {Gerin}, M. 1985, \aap, 150, 327

\bibitem[{{Do} {et~al.}(2020){Do}, {David Martinez}, {Kerzendorf},
  {Feldmeier-Krause}, {Arca Sedda}, {Neumayer}, \& {Gualandris}}]{Do2020}
{Do}, T., {David Martinez}, G., {Kerzendorf}, W., {et~al.} 2020, \apjl, 901,
  L28

\bibitem[{{Do} {et~al.}(2018){Do}, {Kerzendorf}, {Konopacky}, {Marcinik},
  {Ghez}, {Lu}, \& {Morris}}]{Do2018}
{Do}, T., {Kerzendorf}, W., {Konopacky}, Q., {et~al.} 2018, \apjl, 855, L5

\bibitem[{{Do} {et~al.}(2015){Do}, {Kerzendorf}, {Winsor}, {St{\o}stad},
  {Morris}, {Lu}, \& {Ghez}}]{do2015}
{Do}, T., {Kerzendorf}, W., {Winsor}, N., {et~al.} 2015, \apj, 809, 143

\bibitem[{{Eliche-Moral} {et~al.}(2011){Eliche-Moral},
  {Gonz{\'a}lez-Garc{\'\i}a}, {Balcells}, {Aguerri}, {Gallego}, {Zamorano}, \&
  {Prieto}}]{Eliche-Moral2011}
{Eliche-Moral}, M.~C., {Gonz{\'a}lez-Garc{\'\i}a}, A.~C., {Balcells}, M.,
  {et~al.} 2011, \aap, 533, A104

\bibitem[{{Feldmeier-Krause} {et~al.}(2020){Feldmeier-Krause}, {Kerzendorf},
  {Do}, {Nogueras-Lara}, {Neumayer}, {Walcher}, {Seth}, {Sch{\"o}del}, {de
  Zeeuw}, {Hilker}, {L{\"u}tzgendorf}, {Kuntschner}, \&
  {Kissler-Patig}}]{Feldmeier-Krause2020}
{Feldmeier-Krause}, A., {Kerzendorf}, W., {Do}, T., {et~al.} 2020, \mnras

\bibitem[{{Feldmeier-Krause} {et~al.}(2017){Feldmeier-Krause}, {Kerzendorf},
  {Neumayer}, {Sch{\"o}del}, {Nogueras-Lara}, {Do}, {de Zeeuw}, \&
  {Kuntschner}}]{Feldmeier-Krause2017}
{Feldmeier-Krause}, A., {Kerzendorf}, W., {Neumayer}, N., {et~al.} 2017,
  \mnras, 464, 194

\bibitem[{{Figer} {et~al.}(2004){Figer}, {Rich}, {Kim}, {Morris}, \&
  {Serabyn}}]{Figer2004}
{Figer}, D.~F., {Rich}, R.~M., {Kim}, S.~S., {Morris}, M., \& {Serabyn}, E.
  2004, \apj, 601, 319

\bibitem[{{Fragkoudi} {et~al.}(2016){Fragkoudi}, {Athanassoula}, \&
  {Bosma}}]{Fragkoudi2016}
{Fragkoudi}, F., {Athanassoula}, E., \& {Bosma}, A. 2016, \mnras, 462, L41

\bibitem[{{Fritz} {et~al.}(2020){Fritz}, {Patrick}, {Feldemier-Krause},
  {Sch\'odel}, {Schultheis}, {Gerhard}, {Nandakumar}, {Neumayer},
  {Nogueras-Lara}, \& {Prieto}}]{fritz2020}
{Fritz}, T.~K., {Patrick}, L., {Feldemier-Krause}, A., {et~al.} 2020, arXiv
  e-prints, arXiv:3494084

\bibitem[{{Fux}(1999)}]{Fux1999}
{Fux}, R. 1999, \aap, 345, 787

\bibitem[{{Gadotti} {et~al.}(2020){Gadotti}, {Bittner}, {Falc{\'o}n-Barroso},
  {M{\'e}ndez-Abreu}, {Kim}, {Fragkoudi}, {de Lorenzo-C{\'a}ceres}, {Leaman},
  {Neumann}, {Querejeta}, {S{\'a}nchez-Bl{\'a}zquez}, {Martig},
  {Mart{\'\i}n-Navarro}, {P{\'e}rez}, {Seidel}, \& {van de Ven}}]{Gadotti2020}
{Gadotti}, D.~A., {Bittner}, A., {Falc{\'o}n-Barroso}, J., {et~al.} 2020, \aap,
  643, A14

\bibitem[{{Gadotti} {et~al.}(2019){Gadotti}, {S{\'a}nchez-Bl{\'a}zquez},
  {Falc{\'o}n-Barroso}, {Husemann}, {Seidel}, {P{\'e}rez}, {de
  Lorenzo-C{\'a}ceres}, {Martinez-Valpuesta}, {Fragkoudi}, {Leung}, {van de
  Ven}, {Leaman}, {Coelho}, {Martig}, {Kim}, {Neumann}, \&
  {Querejeta}}]{Gadotti2019}
{Gadotti}, D.~A., {S{\'a}nchez-Bl{\'a}zquez}, P., {Falc{\'o}n-Barroso}, J.,
  {et~al.} 2019, \mnras, 482, 506

\bibitem[{{Gallego-Cano} {et~al.}(2020){Gallego-Cano}, {Sch{\"o}del},
  {Nogueras-Lara}, {Dong}, {Shahzamanian}, {Fritz}, {Gallego-Calvente}, \&
  {Neumayer}}]{Gallego-Cano2020}
{Gallego-Cano}, E., {Sch{\"o}del}, R., {Nogueras-Lara}, F., {et~al.} 2020,
  \aap, 634, A71

\bibitem[{{Ginsburg} {et~al.}(2016){Ginsburg}, {Henkel}, {Ao}, {Riquelme},
  {Kauffmann}, {Pillai}, {Mills}, {Requena-Torres}, {Immer}, {Testi}, {Ott},
  {Bally}, {Battersby}, {Darling}, {Aalto}, {Stanke}, {Kendrew}, {Kruijssen},
  {Longmore}, {Dale}, {Guesten}, \& {Menten}}]{Ginsburg2016}
{Ginsburg}, A., {Henkel}, C., {Ao}, Y., {et~al.} 2016, \aap, 586, A50

\bibitem[{{Gonzalez} {et~al.}(2012){Gonzalez}, {Rejkuba}, {Zoccali}, {Valenti},
  {Minniti}, {Schultheis}, {Tobar}, \& {Chen}}]{gonzalez2012}
{Gonzalez}, O.~A., {Rejkuba}, M., {Zoccali}, M., {et~al.} 2012, \aap, 543, A13

\bibitem[{{Kauffmann} {et~al.}(2017){Kauffmann}, {Pillai}, {Zhang}, {Menten},
  {Goldsmith}, {Lu}, \& {Guzm{\'a}n}}]{Kauffmann2017}
{Kauffmann}, J., {Pillai}, T., {Zhang}, Q., {et~al.} 2017, \aap, 603, A89

\bibitem[{{Kim} {et~al.}(2011){Kim}, {Saitoh}, {Jeon}, {Figer}, {Merritt}, \&
  {Wada}}]{Kim2011}
{Kim}, S.~S., {Saitoh}, T.~R., {Jeon}, M., {et~al.} 2011, \apjl, 735, L11

\bibitem[{{Kunder} {et~al.}(2020){Kunder}, {P{\'e}rez-Villegas}, {Rich},
  {Ogata}, {Murari}, {Boren}, {Johnson}, {Nataf}, {Walker}, {Bono}, {Koch},
  {Propris}, {Storm}, \& {Wojno}}]{Kunder2020}
{Kunder}, A., {P{\'e}rez-Villegas}, A., {Rich}, R.~M., {et~al.} 2020, \aj, 159,
  270

\bibitem[{{Launhardt} {et~al.}(2002){Launhardt}, {Zylka}, \&
  {Mezger}}]{Launhardt2002}
{Launhardt}, R., {Zylka}, R., \& {Mezger}, P.~G. 2002, \aap, 384, 112

\bibitem[{{Li} {et~al.}(2015){Li}, {Shen}, \& {Kim}}]{Li2015}
{Li}, Z., {Shen}, J., \& {Kim}, W.-T. 2015, \apj, 806, 150

\bibitem[{{Mackereth} {et~al.}(2019){Mackereth}, {Bovy}, {Leung}, {Schiavon},
  {Trick}, {Chaplin}, {Cunha}, {Feuillet}, {Majewski}, {Martig}, {Miglio},
  {Nidever}, {Pinsonneault}, {Aguirre}, {Sobeck}, {Tayar}, \&
  {Zasowski}}]{Ted2019}
{Mackereth}, J.~T., {Bovy}, J., {Leung}, H.~W., {et~al.} 2019, \mnras, 489, 176

\bibitem[{{Maeda} {et~al.}(2002){Maeda}, {Baganoff}, {Feigelson}, {Morris},
  {Bautz}, {Brandt}, {Burrows}, {Doty}, {Garmire}, {Pravdo}, {Ricker}, \&
  {Townsley}}]{Maeda2002}
{Maeda}, Y., {Baganoff}, F.~K., {Feigelson}, E.~D., {et~al.} 2002, \apj, 570,
  671

\bibitem[{{Martin} {et~al.}(2004){Martin}, {Walsh}, {Xiao}, {Lane}, {Walker},
  \& {Stark}}]{Martin2004}
{Martin}, C.~L., {Walsh}, W.~M., {Xiao}, K., {et~al.} 2004, \apjs, 150, 239

\bibitem[{{Mayer} {et~al.}(2008){Mayer}, {Kazantzidis}, \&
  {Escala}}]{Mayer2008}
{Mayer}, L., {Kazantzidis}, S., \& {Escala}, A. 2008, \memsai, 79, 1284

\bibitem[{{Messineo} {et~al.}(2005){Messineo}, {Habing}, {Menten}, {Omont},
  {Sjouwerman}, \& {Bertoldi}}]{Messineo2005}
{Messineo}, M., {Habing}, H.~J., {Menten}, K.~M., {et~al.} 2005, \aap, 435, 575

\bibitem[{{Molloy} {et~al.}(2015){Molloy}, {Smith}, {Evans}, \&
  {Shen}}]{Molloy2015}
{Molloy}, M., {Smith}, M.~C., {Evans}, N.~W., \& {Shen}, J. 2015, \apj, 812,
  146

\bibitem[{{Morelli} {et~al.}(2004){Morelli}, {Halliday}, {Corsini}, {Pizzella},
  {Thomas}, {Saglia}, {Davies}, {Bender}, {Birkinshaw}, \&
  {Bertola}}]{Morelli2004}
{Morelli}, L., {Halliday}, C., {Corsini}, E.~M., {et~al.} 2004, \mnras, 354,
  753

\bibitem[{{Nagayama} {et~al.}(2003){Nagayama}, {Nagashima}, {Nakajima},
  {Nagata}, {Sato}, {Nakaya}, {Yamamuro}, {Sugitani}, \&
  {Tamura}}]{Nagayama2003}
{Nagayama}, T., {Nagashima}, C., {Nakajima}, Y., {et~al.} 2003, in Society of
  Photo-Optical Instrumentation Engineers (SPIE) Conference Series, Vol. 4841,
  Instrument Design and Performance for Optical/Infrared Ground-based
  Telescopes, ed. M.~{Iye} \& A.~F.~M. {Moorwood}, 459--464

\bibitem[{{Ness} {et~al.}(2013){Ness}, {Freeman}, {Athanassoula},
  {Wylie-de-Boer}, {Bland-Hawthorn}, {Asplund}, {Lewis}, {Yong}, {Lane}, \&
  {Kiss}}]{Ness2013}
{Ness}, M., {Freeman}, K., {Athanassoula}, E., {et~al.} 2013, \mnras, 430, 836

\bibitem[{{Neumayer} {et~al.}(2020){Neumayer}, {Seth}, \&
  {B{\"o}ker}}]{Neumayer2020}
{Neumayer}, N., {Seth}, A., \& {B{\"o}ker}, T. 2020, \aapr, 28, 4

\bibitem[{{Nishiyama} {et~al.}(2006){Nishiyama}, {Nagata}, {Kusakabe},
  {Matsunaga}, {Naoi}, {Kato}, {Nagashima}, {Sugitani}, {Tamura}, {Tanab{\'e}},
  \& {Sato}}]{Nishiyama2006}
{Nishiyama}, S., {Nagata}, T., {Kusakabe}, N., {et~al.} 2006, \apj, 638, 839

\bibitem[{{Nishiyama} {et~al.}(2013){Nishiyama}, {Yasui}, {Nagata},
  {Yoshikawa}, {Uchiyama}, {Sch{\"o}del}, {Hatano}, {Sato}, {Sugitani},
  {Suenaga}, {Kwon}, \& {Tamura}}]{nishiyama13}
{Nishiyama}, S., {Yasui}, K., {Nagata}, T., {et~al.} 2013, \apjl, 769, L28

\bibitem[{{Nogueras-Lara} {et~al.}(2018{\natexlab{a}}){Nogueras-Lara},
  {Gallego-Calvente}, {Dong}, {Gallego-Cano}, {Girard}, {Hilker}, {de Zeeuw},
  {Feldmeier-Krause}, {Nishiyama}, {Najarro}, {Neumayer}, \&
  {Sch{\"o}del}}]{Nogueras2018}
{Nogueras-Lara}, F., {Gallego-Calvente}, A.~T., {Dong}, H., {et~al.}
  2018{\natexlab{a}}, \aap, 610, A83

\bibitem[{{Nogueras-Lara} {et~al.}(2018{\natexlab{b}}){Nogueras-Lara},
  {Sch{\"o}del}, {Dong}, {Najarro}, {Gallego-Calvente}, {Hilker},
  {Gallego-Cano}, {Nishiyama}, {Neumayer}, {Feldmeier-Krause}, {Girard},
  {Cassisi}, \& {Pietrinferni}}]{Paco2018}
{Nogueras-Lara}, F., {Sch{\"o}del}, R., {Dong}, H., {et~al.}
  2018{\natexlab{b}}, \aap, 620, A83

\bibitem[{{Nogueras-Lara} {et~al.}(2019){Nogueras-Lara}, {Sch{\"o}del},
  {Gallego-Calvente}, {Dong}, {Gallego-Cano}, {Shahzamanian}, {Girard},
  {Nishiyama}, {Najarro}, \& {Neumayer}}]{Paco2019a}
{Nogueras-Lara}, F., {Sch{\"o}del}, R., {Gallego-Calvente}, A.~T., {et~al.}
  2019, \aap, 631, A20

\bibitem[{{Nogueras-Lara} {et~al.}(2020{\natexlab{a}}){Nogueras-Lara},
  {Sch{\"o}del}, {Gallego-Calvente}, {Gallego-Cano}, {Shahzamanian}, {Dong},
  {Neumayer}, {Hilker}, {Najarro}, {Nishiyama}, {Feldmeier-Krause}, {Girard},
  \& {Cassisi}}]{nogueras2020}
{Nogueras-Lara}, F., {Sch{\"o}del}, R., {Gallego-Calvente}, A.~T., {et~al.}
  2020{\natexlab{a}}, Nature Astronomy, 4, 377

\bibitem[{{Nogueras-Lara} {et~al.}(2020{\natexlab{b}}){Nogueras-Lara},
  {Sch{\"o}del}, {Neumayer}, {Gallego-Cano}, {Shahzamanian},
  {Gallego-Calvente}, \& {Najarro}}]{nogueras2020a}
{Nogueras-Lara}, F., {Sch{\"o}del}, R., {Neumayer}, N., {et~al.}
  2020{\natexlab{b}}, \aap, 641, A141

\bibitem[{{Pizzella} {et~al.}(2002){Pizzella}, {Corsini}, {Morelli}, {Sarzi},
  {Scarlata}, {Stiavelli}, \& {Bertola}}]{Pizzella2002}
{Pizzella}, A., {Corsini}, E.~M., {Morelli}, L., {et~al.} 2002, \apj, 573, 131

\bibitem[{{Rich} {et~al.}(2007){Rich}, {Origlia}, \& {Valenti}}]{Rich2007}
{Rich}, R.~M., {Origlia}, L., \& {Valenti}, E. 2007, \apjl, 665, L119

\bibitem[{{Rich} {et~al.}(2017){Rich}, {Ryde}, {Thorsbro}, {Fritz},
  {Schultheis}, {Origlia}, \& {J{\"o}nsson}}]{rich:17}
{Rich}, R.~M., {Ryde}, N., {Thorsbro}, B., {et~al.} 2017, \aj, 154, 239

\bibitem[{{Ridley} {et~al.}(2017){Ridley}, {Sormani}, {Tre{\ss}}, {Magorrian},
  \& {Klessen}}]{Ridley2017}
{Ridley}, M. G.~L., {Sormani}, M.~C., {Tre{\ss}}, R.~G., {Magorrian}, J., \&
  {Klessen}, R.~S. 2017, \mnras, 469, 2251

\bibitem[{{Rojas-Arriagada} {et~al.}(2017){Rojas-Arriagada}, {Recio-Blanco},
  {de Laverny}, {Mikolaitis}, {Matteucci}, {Spitoni}, {Schultheis}, {Hayden},
  {Hill}, {Zoccali}, {Minniti}, {Gonzalez}, {Gilmore}, {Rand ich}, {Feltzing},
  {Alfaro}, {Babusiaux}, {Bensby}, {Bragaglia}, {Flaccomio}, {Koposov},
  {Pancino}, {Bayo}, {Carraro}, {Casey}, {Costado}, {Damiani}, {Donati},
  {Franciosini}, {Hourihane}, {Jofr{\'e}}, {Lardo}, {Lewis}, {Lind}, {Magrini},
  {Morbidelli}, {Sacco}, {Worley}, \& {Zaggia}}]{Alvaro2017}
{Rojas-Arriagada}, A., {Recio-Blanco}, A., {de Laverny}, P., {et~al.} 2017,
  \aap, 601, A140

\bibitem[{{Rojas-Arriagada} {et~al.}(2020){Rojas-Arriagada}, {Zasowski},
  {Schultheis}, {Zoccali}, {Hasselquist}, {Chiappini}, {Cohen}, {Cunha},
  {Fern{\'a}ndez-Trincado}, {Fragkoudi}, {Garc{\'\i}a-Hern{\'a}ndez},
  {Geisler}, {Gran}, {Lian}, {Majewski}, {Minniti}, {Monachesi}, {Nitschelm},
  \& {Queiroz}}]{alvaro2020}
{Rojas-Arriagada}, A., {Zasowski}, G., {Schultheis}, M., {et~al.} 2020, \mnras,
  499, 1037

\bibitem[{{Sch{\"o}del} {et~al.}(2014){Sch{\"o}del}, {Feldmeier}, {Kunneriath},
  {Stolovy}, {Neumayer}, {Amaro-Seoane}, \& {Nishiyama}}]{Schoedel2014}
{Sch{\"o}del}, R., {Feldmeier}, A., {Kunneriath}, D., {et~al.} 2014, \aap, 566,
  A47

\bibitem[{{Sch{\"o}del} {et~al.}(2020){Sch{\"o}del}, {Nogueras-Lara},
  {Gallego-Cano}, {Shahzamanian}, {Gallego-Calvente}, \&
  {Gardini}}]{Schoedel2020}
{Sch{\"o}del}, R., {Nogueras-Lara}, F., {Gallego-Cano}, E., {et~al.} 2020,
  \aap, 641, A102

\bibitem[{{Sch{\"o}nrich} {et~al.}(2015){Sch{\"o}nrich}, {Aumer}, \&
  {Sale}}]{schoenrich2015}
{Sch{\"o}nrich}, R., {Aumer}, M., \& {Sale}, S.~E. 2015, \apjl, 812, L21

\bibitem[{{Schultheis} {et~al.}(2019){Schultheis}, {Rich}, {Origlia}, {Ryde},
  {Nandakumar}, {Thorsbro}, \& {Neumayer}}]{schultheis2019}
{Schultheis}, M., {Rich}, R.~M., {Origlia}, L., {et~al.} 2019, \aap, 627, A152

\bibitem[{{Schultheis} {et~al.}(2020){Schultheis}, {Rojas-Arriagada}, {Cunha},
  {Zoccali}, {Chiappini}, {Zasowski}, {Queiroz}, {Minniti}, {Fritz},
  {Garc{\'\i}a-Hern{\'a}ndez}, {Nitschelm}, {Zamora}, {Hasselquist},
  {Fern{\'a}ndez-Trincado}, \& {Munoz}}]{Schultheis2020}
{Schultheis}, M., {Rojas-Arriagada}, A., {Cunha}, K., {et~al.} 2020, \aap, 642,
  A81

\bibitem[{{Schultheis} {et~al.}(2009){Schultheis}, {Sellgren},
  {Ram{\'{\i}}rez}, {Stolovy}, {Ganesh}, {Glass}, \&
  {Girardi}}]{schultheis2009}
{Schultheis}, M., {Sellgren}, K., {Ram{\'{\i}}rez}, S., {et~al.} 2009, \aap,
  495, 157

\bibitem[{{Shields} \& {Ferland}(1994)}]{Shields1994}
{Shields}, J.~C. \& {Ferland}, G.~J. 1994, \apj, 430, 236

\bibitem[{{Simmons} {et~al.}(2014){Simmons}, {Melvin}, {Lintott}, {Masters},
  {Willett}, {Keel}, {Smethurst}, {Cheung}, {Nichol}, {Schawinski},
  {Rutkowski}, {Kartaltepe}, {Bell}, {Casteels}, {Conselice}, {Almaini},
  {Ferguson}, {Fortson}, {Hartley}, {Kocevski}, {Koekemoer}, {McIntosh},
  {Mortlock}, {Newman}, {Ownsworth}, {Bamford}, {Dahlen}, {Faber},
  {Finkelstein}, {Fontana}, {Galametz}, {Grogin}, {Gr{\"u}tzbauch}, {Guo},
  {H{\"a}u{\ss}ler}, {Jek}, {Kaviraj}, {Lucas}, {Peth}, {Salvato}, {Wiklind},
  \& {Wuyts}}]{simmons2014}
{Simmons}, B.~D., {Melvin}, T., {Lintott}, C., {et~al.} 2014, \mnras, 445, 3466

\bibitem[{{Sormani} \& {Barnes}(2019)}]{Sormani2019a}
{Sormani}, M.~C. \& {Barnes}, A.~T. 2019, \mnras, 484, 1213

\bibitem[{{Sormani} {et~al.}(2015){Sormani}, {Binney}, \&
  {Magorrian}}]{Sormani2015}
{Sormani}, M.~C., {Binney}, J., \& {Magorrian}, J. 2015, \mnras, 449, 2421

\bibitem[{{Sormani} {et~al.}(2020){Sormani}, {Magorrian}, {Nogueras-Lara},
  {Neumayer}, {Sch{\"o}nrich}, {Klessen}, \&
  {Mastrobuono-Battisti}}]{Sormani2020}
{Sormani}, M.~C., {Magorrian}, J., {Nogueras-Lara}, F., {et~al.} 2020, \mnras,
  499, 7

\bibitem[{{Sormani} {et~al.}(2018{\natexlab{a}}){Sormani}, {Sobacchi},
  {Fragkoudi}, {Ridley}, {Tre{\ss}}, {Glover}, \& {Klessen}}]{Sormani2018b}
{Sormani}, M.~C., {Sobacchi}, E., {Fragkoudi}, F., {et~al.} 2018{\natexlab{a}},
  \mnras, 481, 2

\bibitem[{{Sormani} {et~al.}(2018{\natexlab{b}}){Sormani}, {Tre{\ss}},
  {Ridley}, {Glover}, {Klessen}, {Binney}, {Magorrian}, \&
  {Smith}}]{Sormani2018}
{Sormani}, M.~C., {Tre{\ss}}, R.~G., {Ridley}, M., {et~al.} 2018{\natexlab{b}},
  \mnras, 475, 2383

\bibitem[{{Thorsbro} {et~al.}(2020){Thorsbro}, {Ryde}, {Rich}, {Schultheis},
  {Renaud}, {Spitoni}, {Fritz}, {Mastrobuono-Battisti}, {Origlia}, {Matteucci},
  \& {Sch{\"o}del}}]{Thorsbro2020}
{Thorsbro}, B., {Ryde}, N., {Rich}, R.~M., {et~al.} 2020, \apj, 894, 26

\bibitem[{{Tress} {et~al.}(2020){Tress}, {Sormani}, {Glover}, {Klessen},
  {Battersby}, {Clark}, {Hatchfield}, \& {Smith}}]{Tress2020}
{Tress}, R.~G., {Sormani}, M.~C., {Glover}, S. C.~O., {et~al.} 2020, \mnras,
  499, 4455

\bibitem[{{Tsatsi} {et~al.}(2017){Tsatsi}, {Mastrobuono-Battisti}, {van de
  Ven}, {Perets}, {Bianchini}, \& {Neumayer}}]{Tsatsi2017}
{Tsatsi}, A., {Mastrobuono-Battisti}, A., {van de Ven}, G., {et~al.} 2017,
  \mnras, 464, 3720

\bibitem[{{Zasowski} {et~al.}(2017){Zasowski}, {Cohen}, {Chojnowski},
  {Santana}, {Oelkers}, {Andrews}, {Beaton}, {Bender}, {Bird}, {Bovy},
  {Carlberg}, {Covey}, {Cunha}, {Dell'Agli}, {Fleming}, {Frinchaboy},
  {Garc{\'\i}a-Hern{\'a}ndez}, {Harding}, {Holtzman}, {Johnson}, {Kollmeier},
  {Majewski}, {M{\'e}sz{\'a}ros}, {Munn}, {Mu{\~n}oz}, {Ness}, {Nidever},
  {Poleski}, {Rom{\'a}n-Z{\'u}{\~n}iga}, {Shetrone}, {Simon}, {Smith},
  {Sobeck}, {Stringfellow}, {Szigeti{\'a}ros}, {Tayar}, \&
  {Troup}}]{zasowski2017}
{Zasowski}, G., {Cohen}, R.~E., {Chojnowski}, S.~D., {et~al.} 2017, \aj, 154,
  198

\bibitem[{{Zasowski} {et~al.}(2013){Zasowski}, {Johnson}, {Frinchaboy},
  {Majewski}, {Nidever}, {Rocha Pinto}, {Girardi}, {Andrews}, {Chojnowski},
  {Cudworth}, {Jackson}, {Munn}, {Skrutskie}, {Beaton}, {Blake}, {Covey},
  {Deshpande}, {Epstein}, {Fabbian}, {Fleming}, {Garcia Hernandez}, {Herrero},
  {Mahadevan}, {M{\'e}sz{\'a}ros}, {Schultheis}, {Sellgren}, {Terrien}, {van
  Saders}, {Allende Prieto}, {Bizyaev}, {Burton}, {Cunha}, {da Costa},
  {Hasselquist}, {Hearty}, {Holtzman}, {Garc{\'{\i}}a P{\'e}rez}, {Maia},
  {O'Connell}, {O'Donnell}, {Pinsonneault}, {Santiago}, {Schiavon}, {Shetrone},
  {Smith}, \& {Wilson}}]{zasowski2013}
{Zasowski}, G., {Johnson}, J.~A., {Frinchaboy}, P.~M., {et~al.} 2013, \aj, 146,
  81

\bibitem[{{Zasowski} {et~al.}(2016){Zasowski}, {Ness}, {Garc{\'\i}a P{\'e}rez},
  {Martinez-Valpuesta}, {Johnson}, \& {Majewski}}]{Zasowski2016}
{Zasowski}, G., {Ness}, M.~K., {Garc{\'\i}a P{\'e}rez}, A.~E., {et~al.} 2016,
  \apj, 832, 132

\bibitem[{{Zoccali} {et~al.}(2014){Zoccali}, {Gonzalez}, {Vasquez}, {Hill},
  {Rejkuba}, {Valenti}, {Renzini}, {Rojas-Arriagada}, {Martinez-Valpuesta},
  {Babusiaux}, {Brown}, {Minniti}, \& {McWilliam}}]{Zoccali2014}
{Zoccali}, M., {Gonzalez}, O.~A., {Vasquez}, S., {et~al.} 2014, \aap, 562, A66

\bibitem[{{Zoccali} {et~al.}(2017){Zoccali}, {Vasquez}, {Gonzalez}, {Valenti},
  {Rojas-Arriagada}, {Minniti}, {Rejkuba}, {Minniti}, {McWilliam}, {Babusiaux},
  {Hill}, \& {Renzini}}]{zoccali2017}
{Zoccali}, M., {Vasquez}, S., {Gonzalez}, O.~A., {et~al.} 2017, \aap, 599, A12

\end{thebibliography}

\end{document}